# Organic Molecules and Water in the Inner Disks of T Tauri Stars

John S. Carr

Naval Research Laboratory, Code 7211, Washington, DC 20375, USA

carr@nrl.navy.mil

Joan R. Najita

National Optical Astronomy Observatory, 950 N. Cherry Avenue, Tucson, AZ 85716, USA

najita@noao.edu

## ABSTRACT

We report high signal-to-noise *Spitzer* IRS spectra of a sample of eleven classical T Tauri stars. Molecular emission from rotational transitions of $H_2O$ and OH and ro-vibrational bands of simple organic molecules ($CO_2$, HCN, $C_2H_2$) is common among the sources in the sample. The emission shows a range in both flux and line-to-continuum ratio for each molecule and in the flux ratios of different molecular species. The gas temperatures (200–800 K) and emitting areas we derive are consistent with the emission originating in a warm disk atmosphere in the inner planet formation region at radii < 2 AU. The $H_2O$ emission appears to form under a limited range of excitation conditions, as demonstrated by the similarity in relative strengths of $H_2O$ features from star to star and the narrow range in derived temperature and column density. Emission from highly excited rotational levels of OH is present in all stars; the OH emission flux increases with the stellar accretion rate, and the OH/$H_2O$ flux ratio shows a relatively small scatter. We interpret these results as evidence for OH production via FUV photo-dissociation of $H_2O$ in the disk surface layers. No obvious explanation is found for the observed range in the relative emission strengths of different organic molecules or in their strength with respect to water. We put forward the possibility that these variations reflect a diversity in organic abundances due to star-to-star differences in the C/O ratio of the inner disk gas. Stars with the largest HCN/$H_2O$ flux ratios in our sample have the largest disk masses. While larger samples are required to confirm this, we speculate that such a trend could result if higher mass disks are more efficient at planetesimal formation and sequestration of water in the outer disk, leading to enhanced C/O ratios and abundances of organic molecules in the inner disk. A comparison of our derived HCN to $H_2O$ column density ratio to comets, hot cores, and outer T Tauri star disks suggests that the inner disks are chemically active.

## 1. INTRODUCTION

Circumstellar disks around recently formed stars are the birthplaces of planets. Studies of molecules in these disks can provide insights into the planet formation process and the degree of chemical processing in disks. For example, the abundance, distribution, and evolution of water in disks are expected to have important effects on planetary system formation, including the growth of giant planets, the oxidation state of the inner disk, and the origin of water in terrestrial bodies. Protoplanetary disks could also be important sources for compounds of prebiotic interest, but the extent to which disks synthesize or preserve organic molecules is unknown. The cold gaseous component of disks at large disk radii (> 40 AU) has been probed with millimeter spectroscopy (Dutrey et al. 2007). Hot molecular gas at small disk radii, including CO, $H_2O$ and OH, have been studied with near-infrared spectroscopy (see Najita et al. 2007a for a review). The main planet formation region of the disk, however, is found at intermediate radii and temperatures, where gas is best studied at mid- and far-infrared wavelengths. Relatively recent developments have made gas at these radii accessible to study.

Spectroscopy in the mid-infrared with the *Spitzer Space Telescope* has opened a new window on gas in the inner planet formation region of disks by the detection of water and organic molecules. Organic molecules were first detected in absorption. Lahuis et al. (2006a) reported absorption from warm (~300-700 K) HCN, $C_2H_2$, and $CO_2$ in the spectrum of the young star IRS 46. The molecular absorption is considered to originate in the inner few AU of a disk viewed close to edge-on. Such absorption is rare (Lahuis et al. 2006a); additional examples include GV Tau (Gibb et al. 2007, 2008; Doppmann et al. 2008) and DG Tau B (Pontoppidan et al. 2008; Kruger et al. 2010).

Carr & Najita (2008) were the first to discover emission from water, OH and organics (HCN, $C_2H_2$, $CO_2$) in *Spitzer* spectra, as reported for the T Tauri star AA Tau. The temperatures and emitting areas of the detected species were consistent with the emission arising in a temperature inversion region of the inner disk atmosphere within a few AU of the star, i.e., in the inner planet formation region of the disk. The relative column densities indicated a high abundance of water vapor in the inner disk, and comparisons to hot cores, comets, and chemical models suggested that inner disks are chemically active. Unlike disk absorption, which requires a special edge-on viewing geometry, the ability to detect molecules in emission means that organic molecules and water can potentially be studied in a large number of T Tauri disks.

Other examples of mid-infrared molecular emission in T Tauri stars were reported from subsequent analysis of existing *Spitzer* IRS data. Salyk et al. (2008) presented the detection of water and OH emission from two T Tauri disks (DR Tau and AS 205). They also described high-resolution near-infrared spectra of $H_2O$ and OH for the same stars. The measured line velocity widths were consistent with the emission originating within a few AU of the stars, and LTE analysis indicated large columns of warm gas. Pascucci et al. (2009) reported HCN and $C_2H_2$ emission from several T Tauri stars, based on spectra taken in the low-resolution mode of *Spitzer* IRS. They also presented observations of young brown dwarfs and the first detections of $C_2H_2$ emission from their disks.

Studies of larger samples of young stars have fully established that mid-infrared emission from water and other molecules is a common property of T Tauri stars. Pontoppidan et al. (2010a) presented a search for molecular emission in 73 stars. The sample included T Tauri stars in Chamaeleon, Ophiuchus and Lupus that are part of a dedicated *Spitzer* cycle-5 program on molecular emission (PID 50641, J. Carr, PI), re-processed spectra from the c2d Legacy program (Evans et al. 2003), and archival spectra of Herbig Ae/Be stars. Emission from $H_2O$, HCN, $CO_2$, $C_2H_2$, and OH is shown to be widespread among the T Tauri stars. On the other hand, the Herbig Ae/Be stars lack detectable molecular emission, with upper limits on line/continuum ratios 3-10 times lower than measured in T Tauri stars. However, hints of $H_2O$/OH emission are seen in the spectra of some Herbig Ae/Be stars and one shows $CO_2$ emission. Detailed analysis of this sample of stars is presented in Salyk et al. (2011).

In this paper we analyze in detail high-quality *Spitzer* spectra of a sample of eleven T Tauri stars. These stars constitute the initial cycle-2 *Spitzer* program that carried out a deep search for $H_2O$ and other gas emission lines in T Tauri stars, which included the observations of AA Tau reported in Carr & Najita (2008). A simple LTE slab model is used to decompose the spectra into contributions from water and organics and determine the emitting gas properties for each molecular species. We describe the characteristics and discuss the origins of the molecular emission. Ratios of our derived molecular column densities are used as estimates of relative abundances to compare to results for hot cores, comets, outer T Tauri disks and protostellar envelopes, and chemical models of disk atmospheres.

## 2. OBSERVATIONS
### 2.1 Spitzer IRS Spectroscopy

The mid-infrared spectra presented here were obtained with the *Spitzer* IRS (Houck et al. 2004) using the Short-High (10–19 μm) module, with a nominal resolving power of ~ 600. All spectra were obtained in staring mode, which places the target at two nod positions along the length of the 11x4.7" slit. The total integration times were set using the Spectroscopy Performance Estimation Tool to achieve a continuum signal-to-noise ratio (S/N) ≥ 300, using 8 to 24 nod cycles. The redundancy provided by the large number of cycles enabled the evaluation of the noise statistics and identification of badly behaved pixels. The SH frame time and number of cycles for each target are given in Table 1. Dedicated background observations were obtained for all targets at off positions clean of continuum sources or known nebulosity. In most cases, the background integration time was one-half of the time spent on the target.

The T Tauri stars studied in this paper were observed in *Spitzer* GO program 2300. These observations were designed primarily to search for $H_2O$ emission from disks, before such emission was known to be common among T Tauri stars. Results from this program for AA Tau were first reported in Carr & Najita (2008). The program also obtained spectra for all targets in the Long-High (19–37 μm) module; analyses of those data will be presented in a future paper.

The main part of this sample are 9 well-studied classical T Tauri stars in the Taurus-Auriga star forming region, selected to cover a range in mass accretion rate. Spectral types range from K3 to M0. The program included two additional sources, AS353A and V1331 Cyg, both of which are extremely active stars known to have CO overtone, $H_2O$ or OH emission in the near-infrared (Najita et al. 2007a, 2009; Doppmann et al. 2010; Prato et al. 2003). V1331 Cyg is often presumed to have a spectral type and mass intermediate between T Tauri stars and Herbig Ae stars due to the non-detection of photospheric features and a higher luminosity than typical T Tauri stars (e.g., Hamann & Persson 1992). Of the Taurus sources, DG Tau has also shown CO overtone and near-infrared water emission (Carr 1989; Najita et al. 2000), but the CO overtone emission varies and is not always present (Greene & Lada 1996; Biscaya et al. 1997). V1331 Cyg, AS353A, and DG Tau also have known jets and/or outflows.

The star UY Aur is a 0.9" binary whose components have similar accretion rates (Hartigan & Kenyon 2003) and comparable brightness in the infrared; the binary flux ratio at 10 μm has varied from 1.2 to 2.0 over a decade (Skemer et al. 2010). The 10 μm silicate spectra of the components have been spatially resolved by Skemer et al. (2010). However, we cannot determine the contribution of the components to the line emission spectrum in the unresolved *Spitzer* IRS spectra. Two other stars have companions within the SH slit, DK Tau and RW Aur, with separations of 2.4 and 1.4", respectively. The primary dominates the mid-infrared flux in each case, with flux ratios ~ 10 (McCabe et al. 2006).

Some properties of the sample that are used in plots later in this paper are listed in Table 1. The stellar accretion rates for most stars are those reported in Najita et al. (2007b), which placed accretion rates from different literature sources on the same scale as Gullbring et al. (1998). Accretion rates are probably no better than a few tenths of a dex due to intrinsic variability. For GI Tau, we adopted the accretion rate from Hartmann et al. (1998), which is on the same scale as Gullbring et al. We scaled down the accretion rate for AS353A (Hartigan et al. 1995) by 0.12, based on the prescription of Najita et al. (2007b). We also made a rough estimate of the accretion rate of V1331 Cyg based on the accretion flux derived by Eisner et al. (2007). Assuming a distance of 550 pc (as in Najita et al. 2009), the corresponding accretion rate is ~$10^{-6}$ $M_{sun}$ yr$^{-1}$. The disk masses are from the submillimeter measurements in Andrews & Williams (2005).

## 2.2. Data Reduction

The stars in the sample are sufficiently bright that systematic effects become the limiting factor in the achievable S/N. Therefore, the data processing procedures are critical in maximizing the quality of the spectra. Our procedures were initially described in Carr & Najita (2008), and we cover these in more detail here. The data processing was carried out using a combination of standard IRAF tasks and custom routines.

We begin with the non-flat-fielded "droop" images, multiplied by the frame time. For each target, frames are combined to form a mean image for the "A" slit position, the "B" slit position, and the combined background positions. When combining frames, the noise is calculated for each pixel and any highly deviant values are rejected. In addition, pixels in the background image that have a standard deviation much higher than the distribution for all pixels are flagged as being unstable; these are similar to the original definition for "rogue pixels" for the IRS arrays. These flagged pixels are used to make an unstable pixel mask, which is combined with known permanently dead pixels for a total bad pixel mask. The mean background image is then subtracted from the mean A and B images, which removes any background signal, dark current, and hot (but stable) pixels. These images are next divided by a flat-field. We use the pipeline flat-field that has been normalized in both the spectral and spatial dimensions using the IRAF task apflatten. This avoids artificial increases in the noise where the flat drops quickly near the ends of the slit and preserves the noise statistics. After division by the flat-field, the bad pixel mask is applied, replacing bad pixels by interpolation in the spectral direction. A visual inspection is also carried out and any missed and obviously bad pixels are fixed by hand. The multi-order spectra are traced and extracted with optimal weighting using the IRAF task apall, making use of the noise statistics determined from the data. The wavelength scale for the extracted spectra was obtained by fitting a polynomial function to the pipeline wavelength calibration table.

The extracted orders were divided by a spectral response function to remove the instrumental spectral shape and fringing and to flux calibrate the spectra. The spectral response function was customized for each slit position of each target using a suite of spectra for the IRS standard stars $\xi$ Dra and $\delta$ Dra. These data were obtained from the *Spitzer* archive and processed in the same way as the target spectra, with an additional step of dividing the spectra by a corresponding model template provided on the *Spitzer* IRS website. The quality of fringe correction is very sensitive to the choice of spectrum in the suite of calibrators; most likely this is due to small differences in centering of stars in the slit. We divided the target spectrum by each of the calibrator spectra to find those that minimized the residual fringing. For the SH module, we examined order 20, which usually shows the largest residual fringing and showed few intrinsic emission features in the T Tauri stars. Those standard star spectra (typically 1–4) that gave the smallest residual fringing for a target and slit position were averaged for the spectral response function. Standard star spectra that minimized the residual fringing also eliminated significant curvature in the divided spectra at the ends of the orders.

While this procedure provides a good correction for fringing in our sample of stars, there is often some residual fringing seen in the highest (shortest wavelength) orders. To correct for this remaining residual we used the routine IRSFRINGE (Lahuis et al. 2006b) to search for and remove any significant fringes near the main fringe frequencies of 3500 and 7800 cycles. However, no correction was applied for the 3500 cycle fringe in orders near the HCN and $C_2H_2$ bands, because the R and P-branch spacing for these molecules is close to this frequency.

Finally, the individual spectral orders were combined using appropriate weighting in the wavelength overlap regions. No scaling of individual orders was required because the match in flux levels between adjacent orders was excellent. The A and B spectra were then averaged. A comparison of the A and B spectra was used to estimate the noise in the mean spectrum. The continuum S/N at 14 μm ranged from 200 to 400.

### 3. OBSERVED SPECTRA

The complete Short-High spectra are presented in Figure 1. The low-frequency variations in the continuum are due to dust emission features, in particular the amorphous silicate bands centered at 10 and 18 μm. Crystalline silicates are present in many objects, with forsterite emission centered at 11.2 μm, 16 μm, and 11.9 μm, and silica emission near 12.5 μm. We also find weak and shallow emission features centered roughly at 13.7, 14.5 and (less discernible) 15.5 μm that sometimes underlay the forest of emission lines. These peaks correspond in wavelength with pyroxenes (Jager et al. 1998; Chihara et al. 2002), but curiously the corresponding peaks of the 10 μm complex are not apparent in the spectra. No broadband absorption features are identified in the spectra.

As was described by Carr & Najita (2008), rotational lines of $H_2O$ dominate the forest of emission lines seen above the dust continuum. Some of the rotational transitions that make up the water emission features are identified in Figure 4 of Pontoppidan et al. (2010a). Relative to the continuum, the water emission ranges from very prominent (e.g., RW Aur) to barely discernable (e.g., DO Tau). Water emission is not detected in three stars: DG Tau, V1331 Cyg and AS 353A. The other molecules contributing to the emission spectrum are rotational transitions of OH and the ro-vibrational fundamental bending modes of HCN ($v_2$), $C_2H_2$ ($v_5$) and $CO_2$ ($v_2$), with Q-branches at 14.0, 13.7, and 14.97 μm, respectively.

A continuum was subtracted from each spectrum in order to facilitate modeling and measurement of the emission features. The definition of the continuum is challenging because there are few true continuum regions due to the density of emission lines and the low spectral resolution. This is further complicated by the presence of the dust emission features, which increases the uncertainty in the continuum level for water lines near 12.5, 14.5, and 16 μm. In addition, the HCN and $C_2H_2$ bands overlay the wide and shallow 13.7 μm dust feature. We used our spectral modeling (§4) to provide a guide to the best continuum or pseudo-continuum regions. Polynomials were fit for 6 to 8 separate intervals over the SH range, and these individual continuum subtracted regions were joined into a final spectrum. All the continuum subtracted spectra are presented in Figure 2 for the 12-16 μm region, which contains the organic molecular bands.

The complexity of the molecular spectrum in this region is demonstrated in Figure 3, which shows synthetic emission spectra for the individual molecules calculated using the methods described in §4. Figure 3(a) shows that there is significant blending of the molecular emission from different species at the resolution of the IRS SH module. Failure to account for this blending can lead to erroneous fluxes or false detections for a molecule, depending on the relative emission strength of different species. Furthermore, nearly all observed molecular features are blends of individual rotational or ro-vibrational transitions (compare to Figure 3(b), the same model at high-resolution).

Emission from OH is detected from every star in the sample, including those in which $H_2O$ emission is not detected. The most isolated and easily measured OH feature is at 14.64 μm (see Fig. 2). This feature is a blend of unresolved components of the ground vibrational state with total angular momentum quantum number N=19. Other rotational OH emission features, ranging from N = 14 to as high as N = 29, can be seen in the spectra over the SH spectral range (see Najita et al. 2010 for a list of OH wavelengths in this spectral region).

Prominent HCN emission is clearly present in 5 stars of the sample (including the previously reported case of AA Tau). HCN emission may be present in UY Aur, but our modeling (§4) shows that a large part of the 14.0 μm feature in UY Aur is due to $H_2O$ and OH. $C_2H_2$ emission is present in at least 3 of these stars, but modeling is required to account for the $H_2O$ and HCN emission in the 13.7 μm feature in others.

The most commonly observed organic molecule is $CO_2$. It is detected in all stars except DG Tau, with a questionable detection in DO Tau. $CO_2$ is observed in some stars without HCN or $C_2H_2$ emission and in two that lack $H_2O$ emission (V1331 Cyg and AS 353A). Other stars showing only $CO_2$ emission were reported by Pontoppidan et al. (2010a). Note that a pair of $H_2O$ emission features is blended with the blue wing of the $CO_2$ Q-branch (see Figure 2).

Najita et al. (2010) reported the identification of a Q-branch of $HCO^+$ at 12.06 μm in the *Spitzer* spectrum of the transitional disk object TW Hya. We do not detect this $HCO^+$ band in any star in our sample of T Tauri stars. $HCO^+$ emission with the distance-corrected flux measured in TW Hya would likely have escaped detection in our sample of Taurus objects, especially since the $HCO^+$ Q-branch is close to an observed water feature at 12.08 μm. The detection of HCO+ emission from TW Hya was simplified by the notable lack of strong water emission.

One notable result is that the relative intensities of different molecular species differ from object to object. The spectra in Figure 2 show that the flux ratios of organic molecules to water, and the flux ratios of different organic bands, are not constant. Compare, for example, the relative strength of $CO_2$ to HCN in UY Aur and BP Tau, or the relative strength of HCN to water lines near 14.4 μm in AA Tau and GK Tau. We return to this point in §5, where we make use of representative flux measurements for the emission from each molecule.

There are also some atomic emission lines observed in the SH spectra. The [NeII] line at 12.81 μm is very common in this sample, but it is often quite weak and is close in wavelength to a number of weak molecular lines. It appears most prominently in AA Tau, UY Aur, and DG Tau. Atomic hydrogen has a number of transitions in this wavelength region (see Najita et al. 2010), with the strongest line at 12.37 μm line. However, this line is partially blended with $H_2O$ emission features. The 17.93 μm line of [Fe II] is the most prominent emission line in V1331 Cyg and AS353A and is also present in DG Tau. In other stars, this [Fe II] line is coincident with a water emission feature.

## 4. SYNTHETIC SPECTRA

### 4.1 Modeling Method

Following the methods in Carr & Najita (2008), we calculated synthetic spectra of the detected molecular species in order to place constraints on the gas conditions producing the emission. This approach adopts a plane parallel slab model with a single temperature and column density for each molecule and assumes that all level populations are in LTE. The three variables required to fit the spectrum for a particular molecule are the temperature T, the line-of-sight (LOS) column density N of the molecule, and the projected emitting area. We define $R_e$ as the radius of a circular region with this emitting area. The temperature and column density are determined by matching the relative strengths of spectral features (usually blends of lines) in the case of $H_2O$ and OH, or by fitting the shape of the Q-branch in the case of HCN, $C_2H_2$ and $CO_2$. For each combination of temperature and column density, $R_e$ is adjusted to match the observed absolute flux. We neglect continuum dust opacity within the line-emitting layer. The local line width was taken to be due to thermal broadening without assuming any additional turbulent motions. The synthetic spectrum was first calculated at a resolution sufficient to sample the local line width and then smoothed to the resolution of IRS. The result was binned to the same wavelength sampling as the observed spectra to allow easy direct comparison.

The molecular line data are taken from the HITRAN database (Rothman et al. 2004), with the exception of HCN. For HCN we used a theoretical linelist from Harris et al. (2006), which includes higher vibrational bands than are present in the HITRAN linelist. The Harris et al. line strengths must be increased by a factor of 6 to account for the state independent degeneracy factor (Fischer et al. 2003) as used in the HITRAN database and its associated partition functions. Because the calculated line strengths from Harris et al. are somewhat larger than the corresponding ones in HITRAN, we further scaled them by a factor of 0.88 in order to be consistent with synthetic spectra that use the HITRAN data for HCN. For $H_2O$, we also ran models with the BT2 linelist (Barber et al. 2006). We found no significant differences between spectra computed with the BT2 and HITRAN linelists for the range of temperatures that fit the IRS water spectra. However, we caution that the HITRAN linelist will be inadequate at significantly higher (> 1000 K) temperatures. The partition functions were calculated using available FORTRAN routines from the work of Fischer et al. (2003), which are consistent with the definition of statistical weights in the HITRAN database.

### 4.1.1 $H_2O$ and OH

In modeling the water emission, we focused on obtaining the best fit to the $H_2O$ spectrum in the 12–16 μm region. The energy levels for the $H_2O$ transitions have no overall trend with wavelength and the lines in the 12–16 μm interval are representative of the range of upper energy levels within the wavelength coverage of the IRS. In our experience, this approach provides a reasonable characterization of the average water emission properties in the full *Spitzer* spectrum (but see §4.2.2). By focusing on this wavelength region we can also optimize the removal of the water emission spectrum in the region of the organic emission bands, which aids in modeling these features. We also fit and remove an OH emission spectrum for the same purpose.

Synthetic $H_2O$ spectra were calculated over a range in temperature and column density and compared to the observed spectrum. The final $H_2O$ model parameters were based on a best fit by eye to the spectrum. The approach used for the organic bands in §4.1.2, of calculating $\chi^2$ from the difference of the model and observed spectrum, did not work as well over the large wavelength interval for $H_2O$; variations in the zero continuum level and intervals with weak or no lines had undue influence on the fit. Table 2 gives the model parameters adopted for $H_2O$. An example synthetic spectrum is over-plotted on the observed spectrum of BP Tau in Figure 4. For GK Tau and DO Tau, $H_2O$ is detected but the S/N in the emission lines is insufficient to constrain the parameters well. For the purpose of subtracting a synthetic $H_2O$ spectrum from the data for these two stars, scaled synthetic spectra were calculated using the mean temperature and column density in Table 2.

Emission from high rotational transitions of OH are present throughout the SH spectral region in all of the stars in our sample. After subtraction of the synthetic $H_2O$ spectrum, a model OH spectrum was calculated to match the OH spectrum (see Fig. 4 for an example). We subtract this OH synthetic spectrum from the data, though the only feature that has some impact on our results is the OH feature at 14.06 μm (N=20), which blends with the red wing of the HCN band (see Fig. 3).

### 4.1.2 The Organics: HCN, $C_2H_2$ and $CO_2$

In modeling the HCN, $C_2H_2$ and $CO_2$ bands, we first subtract the best $H_2O$ + OH model for each object from the observed spectrum. $H_2O$ can contribute significantly to the flux within the Q-branches of HCN and $C_2H_2$. In addition, both HCN and $C_2H_2$ contribute emission to the Q-branch of the other, which can be important when the emission from one molecule is significantly greater than the other. BP Tau is a good example where this blending produces an apparent $C_2H_2$ feature. As can be seen in Figure 4, BP Tau shows emission near 13.7 μm that could be mistaken for the $C_2H_2$ band; however, the model in Figure 4 shows that $H_2O$ emission accounts for most of this flux. Further modeling shows HCN emission can account for the remaining residual.

The shapes of the Q-branches are determined by both temperature and column density. Synthetic spectra were calculated on a grid in T and N, scaling $R_e$ to match the model flux to the observed flux integrated over the band. $\chi^2$ was calculated at each grid point from the difference of the model and observed spectrum, and contour plots of $\chi^2$ were produced to determine the best-fit values and confidence intervals. There is significant degeneracy between T and N in fitting each of the three molecular bands, with increasing T offset by decreasing N.

When calculating the grid of HCN models, a synthetic $C_2H_2$ spectrum was subtracted from the observed spectrum to account for $C_2H_2$ contribution to the HCN band. This fixed $C_2H_2$ model spectrum was calculated from a combined HCN and $C_2H_2$ model that was close to the final solution based on a first round of fitting. Two examples of contour plots of $\chi^2$ for HCN are shown in Figure 5. The heavy contour is the 90% confidence limit, which we adopt as the uncertainty range.

A similar procedure was followed for fitting the $CO_2$ Q-branch. Before fitting the $CO_2$ band, both the best $H_2O$+OH model and the final HCN+$C_2H_2$ model were subtracted from the data. Example $\chi^2$ contour plots for $CO_2$ are shown in Figure 6.

For $C_2H_2$, we assumed that HCN and $C_2H_2$ have the same excitation temperature and emitting area, and adjusted the $C_2H_2$ abundance relative to HCN to fit the $C_2H_2$ band. The reason for this approach is that T and N were not generally constrained at a useful level of confidence, probably because of the lower flux in the $C_2H_2$ Q-branch and greater contamination from $H_2O$ and HCN. An exception is RW Aur, which has the largest $C_2H_2$ to HCN flux ratio in the sample (see §4.2.1).

### 4.2 Modeling Results
### 4.2.1 Organic Emission

The best-fit (minimum $\chi^2$) synthetic spectra for HCN and $C_2H_2$ are shown in Figure 7, over-plotted on the observed spectra with the best $H_2O$+OH model subtracted. In some cases (e.g., DK Tau), significant residuals remain after fitting the HCN+$C_2H_2$ model; many of these residuals coincide with $H_2O$ features that are not well matched by the best-fit water spectrum. We were able to determine HCN parameters (Table 3) for 5 of the 6 stars in which HCN was detected. For UY Aur, the HCN band is only marginally detected (Fig. 7), but its measured strength is sensitive to the choice of the continuum fit. Emission from $C_2H_2$ in UY Aur can only be considered an upper limit.

The observed HCN Q-branches are a blend of multiple ro-vibrational bands, with significant contribution to the emission from vibrational levels up to v=3 or 4 (depending on the temperature) and rotational levels up to J=24 to 30. In the best cases (AA Tau and BP Tau in Fig. 7), the R-branch lines can be discerned extending to shorter wavelengths to at least the J=22–21 transition. Hence, HCN is both rotationally and vibrationally hot, and the fits in Figure 7 are consistent with a similar rotational and vibrational excitation temperature. A possible exception is GI Tau, which shows an excess at 13.85 μm relative to the model. This could be excess emission from the v=4 and 5 vibrational bands, but an unidentified spectral feature is also a possibility. The definition of the continuum is also an issue at this spectral resolution and wavelength, especially since HCN and $C_2H_2$ sit atop the 13.7 μm solid state feature which has an unknown spectral structure.

The best fit model for the HCN band is always optically thick or marginally optically thick, with τ=0.4 to 4 in the Q(10) line of the v=1-0 band. The HCN column densities for the 5 stars have a range of about 3, with temperatures between 550 and 850 K. The contour plots in Figure 5 show that as the temperature is decreased (or increased) a respective increase (or decrease) in the column density can produce acceptably good fits over a range in parameters. Within the 90% confidence interval, the lower limit on the temperature corresponds to the upper limit on the column density; these parameters are given in Table 3 as the 'T minimum' solution. The upper limits on the HCN column densities are 2-5 times greater than the minimum $\chi^2$ solution.

For each object there also exists a warmer, optically thin solution within the 90% confidence interval. This is where the contours in Figure 5 continue to lower column density at roughly constant temperature. Hence, while the fits place an upper limit on the temperature that corresponds to the optically thin case, there is no formal lower limit on the column density, as long as an arbitrarily large area (but constant number of molecules) can be invoked to produce the total emission flux. A lower limit can be placed on the HCN column by assuming that the emitting area for HCN is no larger than that for $H_2O$. This is a reasonable assumption given that the derived temperatures for HCN are similar to or greater than those found for $H_2O$. This limit, with $R_e(HCN) = R_e(H_2O)$, is given as the 'optically thin' solution in Table 3.

The best-fit HCN solution for AA Tau in Table 3 is very close to that reported in Carr & Najita (2008), although the uncertainty in column density and temperature is much larger in this paper. The reason for this difference is that the uncertainty given by Najita & Carr considered each parameter independently rather than looking at the two-dimensional confidence contours. This also meant that the optically thin solution was not considered as a possibility in Carr & Najita.

As discussed in §4.1.2, the parameters for $C_2H_2$ are not determined independently, but the $C_2H_2$ column density relative to HCN is derived by assuming that $C_2H_2$ and HCN have the same temperature and emitting area. The ratio of the $C_2H_2$ to HCN column densities is given in Table 4 for the same three HCN solutions in Table 3. The $C_2H_2$ column density ranges from 0.04 to 0.4 that of HCN for the best-fit HCN solution.

The assumption that $C_2H_2$ has the same excitation temperature as HCN produces a good fit to the $C_2H_2$ Q-branch in all three cases where the emission is well detected (see Fig. 7), but small temperatures differences (~ 100-200 K) cannot be ruled out. For RW Aur, we can independently fit the $C_2H_2$ band. This gave T=615 K for $C_2H_2$, compared to 690 K for HCN; the $C_2H_2$ column density is 3 times larger than that obtained assuming the same temperature and area. The $\chi^2$ contour plots overlap significantly within their 90 % confidence intervals. For the optically thin solution, the temperature is 715 K for $C_2H_2$ vs. 745 K for HCN. Hence, the modeling uncertainties are consistent with the same temperature for HCN and $C_2H_2$ in RW Aur, though differences in temperature or emitting area would affect the derived relative column densities.

A sampling of best-fit models for the $CO_2$ emission is shown over-plotted on the data in Figure 8. The observed spectra in Figure 8 show differences in the width of the $CO_2$ Q-branch, indicating a range in temperature and/or column density from object to object. However, the modeling is unable to place constraints on the $CO_2$ column density. The examples in Figure 6 show that the column density is not bounded within the 90% confidence contour in either the optically thin or the very optically thick limits. For this reason, we do not report column densities for $CO_2$.

The models can place some constraint on the $CO_2$ excitation temperature. In Table 5, we give the $CO_2$ temperature at the minimum $\chi^2$ solution and the allowable range for the temperature. At the lower temperature end, the 90% confidence contour often extends to temperatures below 75 K (e.g., Fig. 6b), which is the minimum temperature at which the HITRAN partition functions can be applied. We also give the minimum radius for the emitting area that was found within the 90% confidence contour. This radius corresponds to a marginally optically thick solution at the high temperature edge of the confidence area. The average of the best-fit temperatures in Table 5 is 340 K, but there is a considerable range between objects and in the allowable range for each object, and some stars (e.g., BP Tau) are very poorly constrained. The most noteworthy result is that the $CO_2$ temperature is usually lower, often substantially less, than the temperatures derived for $H_2O$ and HCN.

### 4.2.2 $H_2O$ and OH

The single temperature and column density LTE model provides a good first order match to the observed $H_2O$ spectrum, though there are noticeable discrepancies. With the parameters for our best fits in the 12–16 μm region, the largest disagreements are seen for a number of strong, high excitation lines ($E_u > 5000$ K) whose strengths are over predicted by the model, particularly in the 16-18 μm region. Lower temperatures can provide a better fit to these lines, but at the expense of poor fits in the 12-16 μm interval. Meijerink et al. (2009) show in their non-LTE water calculations that transitions with higher energy levels and larger Einstein A-values will be the farthest from LTE. This strongly suggests that we are seeing the effect of sub-thermal populations in many higher rotational levels. Our modeling also shows that the water spectra are dominated by rotational transitions in the ground vibrational state, with only minor contributions from transitions in the first vibrational state. The absence of excess emission from transitions in the first vibrational state, compared to the thermal LTE model, implies that infrared pumping in the fundamental ro-vibrational water bands is not significant.

The spectra in Figure 2 are remarkably similar in the relative strengths of $H_2O$ features from star to star, suggesting similar excitation conditions and optical depth. This impression is confirmed by the best-fit models, which have a relatively narrow range in temperature and column density (Table 2). In fact, within the uncertainties, the results are consistent with a single temperature near 600 K. The $H_2O$ column density shows some variation from star to star, but its range is less than a factor of 5. For each of the objects modeled, the main $H_2O$ lines are optically thick. The narrow parameter range for the $H_2O$ lines is a more robust result than the absolute values for the parameters, because the actual values will be subject to systematic modeling effects, such as assumptions about LTE and local line broadening. Salyk et al. (2011) also find a small parameter range for $H_2O$, but with a somewhat different mean temperature.

Despite the similar temperature of the water emission among the sources in the sample, the emission fluxes vary by a factor of 5 (Table 6). Based on the modeling shown in Table 2, this is primarily the result of the *projected* emitting areas. The range of projected emitting radii is less than a factor of 2, with a mean on the order of 1 AU. Thus, the water emission is likely arising from a similar range of disk radii.

The typical temperature required to match the rotational intensity distribution of the OH lines in the SH spectra was 4000 K. This result is similar to the high rotational temperature found from the OH spectrum of the transition disk TW Hya (Najita et al. 2010). For TW Hya, Najita et al. suggested the possibility of prompt OH emission from the photo-dissociation of $H_2O$. We will discuss this further in §6.3. A detailed analysis of the OH emission in this sample, which combines data from the SH and LH spectra, will be presented in a subsequent paper.

## 5. MOLECULAR FLUXES

### 5.1 Measurement of Fluxes

Fluxes for the molecular and atomic species observed in the SH spectrum were measured from the continuum subtracted spectra. As described in §3, a spectral feature for a given species may contain unresolved contributions from other species, with the transitions of $H_2O$ the most common contaminant. In order to account for this blending, we used the synthetic spectrum modeling described in §4. The fluxes for all species other than $H_2O$ were measured after subtracting the best $H_2O$ model from the spectrum. Synthetic spectra for additional species were subtracted as necessary, as described below. Measured fluxes for the molecular species are given in Table 6, and fluxes for the atomic lines are in Table 7. The uncertainties in the fluxes include both the pixel-to-pixel noise in the spectrum and an uncertainty on the continuum level.

The reported $H_2O$ flux is the combined flux of the three $H_2O$ features centered at 17.09, 17.22 and 17.34 μm. Each of these spectral features is composed of multiple $H_2O$ transitions. The three $H_2O$ features are in close proximity to each other, well separated from other known molecular or atomic species, and among the strongest $H_2O$ features in the SH spectral region. The $H_2O$ flux was measured between 17.075-17.385 μm. For OH, we measured the spectral feature at 14.64 μm. This OH feature is the cleanest in the SH spectrum. It is dominated by a blend of the 4 components of the N=19 transitions of the ground vibrational state of OH.

The Q-branches of HCN and $C_2H_2$ each contain some contribution from transitions of the other. Within this data set, this has a substantial effect only when $C_2H_2$ is weak or undetectable relative to HCN, which is the situation for BP Tau and DK Tau. The HCN flux was measured over the interval from 13.90 to 14.05 μm, after subtracting the best-fit models for $H_2O$, OH and $C_2H_2$. Similarly, the $C_2H_2$ flux was measured over the interval 13.65 to 13.735 μm, after subtraction of the best-fit models for $H_2O$, OH, and HCN. The $CO_2$ Q-branch has a pair of prominent $H_2O$ features that blend with the blue wing of the band. The $CO_2$ flux was measured from 14.93 to 14.997 μm, with the blue edge of the interval chosen to minimize uncertainty from these $H_2O$ features. A combined synthetic model for $H_2O$, OH, HCN and $C_2H_2$ was subtracted before measuring the $CO_2$ flux.

The last molecular feature included in Table 6 is the S(1) transition of molecular hydrogen at 17.03 μm, which sits at a clean location between $H_2O$ lines. The S(2) rotational line at 12.28 μm, on the other hand, is coincident with both $H_2O$ and OH emission features (see Fig. 3). These features normally dominate over any $H_2$ emission in our sample and make it impossible to measure the S(2) line.

Fluxes for atomic lines are given in Table 7. Emission from the [NeII] line at 12.81 μm is present in nearly every object, but it is often not prominent in the spectrum. Because weak features of $H_2O$, OH, HCN and $C_2H_2$ are present near this line, we subtracted the combined synthetic molecular spectrum before measuring the [NeII] flux. The strongest HI line in the spectrum is at 12.37 μm. However, this line is blended with $H_2O$ features that can contribute up to one-half the combined strength of the feature, based on the synthetic $H_2O$ spectrum. This blending could add additional, but unknown, uncertainty to the HI strengths in Table 7. Line fluxes for the [Fe II] line at 17.93 μm are only given for the three stars that lack $H_2O$ emission. For the rest of the objects, the strength of the 17.93 μm water feature predicted by the synthetic spectrum is equal to or greater then the observed strength, and hence it is not possible to extract any information on [Fe II] emission for these stars.

### 5.2 Absolute and Relative Fluxes

As noted in §3, the spectra in Figure 2 show a significant star-to-star variation in the relative fluxes of different molecular species. Figure 9 plots ratios of the measured molecular fluxes from Table 6 against one another. The first two panels show the ratios of HCN, $C_2H_2$, $CO_2$, and OH with respect to $H_2O$. The last panel compares the organic bands among themselves by plotting the ratio of $C_2H_2$ to HCN against $CO_2$ to HCN. These plots clearly establish that there is a real range in the relative fluxes of molecular species, about one order of magnitude in most cases. The range in the ratio $CO_2/H_2O$ is somewhat smaller, a factor of four. The OH to $H_2O$ flux ratio for all but one of the stars falls within a relatively narrow band, about a factor of two in width.

The absolute fluxes for each of the molecular species in Table 6 are also diverse, showing a range of about one order of magnitude for the Taurus sample. There is also an order of magnitude range seen in the line to continuum ratios (or equivalent widths) of the emission features.

## 6. DISCUSSION
### 6.1 Frequency and Origin of the Emission

While there is a substantial range in the fluxes, equivalent widths, and flux ratios of the detected molecular species in our sample, a common element is that mid-infrared molecular emission is ubiquitous. Emission in the high rotational lines of OH is present in all stars. Considering only the Taurus objects, the detection rates for $H_2O$ and $CO_2$ are 89%. Such high detection rates are comparable to the frequency of CO fundamental in classical T Tauri stars (Najita et al. 2003; Salyk et al. 2011). Emission in the Q-branches of HCN and $C_2H_2$ is less common, at 67 and 44 %. These molecular detection rates are comparable to those found in larger samples of classical T Tauri stars (Pontoppidan et al. 2010a).

As we initially reported for the star AA Tau, the analysis of the mid-infrared molecular emission in this larger sample of classical T Tauri stars yields emitting areas and temperatures that are consistent with an origin in the inner few AU of the circumstellar disk. We find that the molecule with the best constrained parameters, $H_2O$, has projected emitting radii of ~ 1 AU, implying de-projected radii < 2 AU. The best-fit values for HCN give smaller emitting areas, 0.1 to 0.4 that of $H_2O$, and projected radii of 0.4–0.6 AU. However, equal areas for HCN and $H_2O$ are also consistent with the errors (§4.2.1). The temperatures for $H_2O$, HCN, and $C_2H_2$ are constrained to be in the range of 400 to 800 K. The temperature for $CO_2$, while often more uncertain, is generally lower and in the range of 100 to 600 K. These temperatures are consistent with models that calculate the vertical thermal structure of the molecular gas at radii within a couple of AU of the star (Glassgold et al. 2004; Nomura et al. 2007; Glassgold et al. 2009; Woitke et al. 2009).

The interpretation of a disk origin for the molecular emission is corroborated by the velocity profiles of $H_2O$ lines that have been measured for a small number of objects with high-resolution mid-infrared spectroscopy (Pontoppidan et al. 2010b; Knez et al. 2007). The line widths are consistent with disk radii of ~ 1 AU and similar to those measured for the CO fundamental transitions. The high critical densities required for thermal excitation of the mid-infrared water transitions (up to $10^{12}$ cm$^{-3}$, Meijerink et al. 2009) provide additional support that the lines form in a disk atmosphere.

Shocks associated with unresolved outflows within the IRS slit are not likely to explain the molecular emission, because the mid-infrared spectra for the T Tauri stars are dissimilar from those observed for outflows. The *Spitzer* IRS spectra of protostellar outflows are generally dominated by emission in the rotational transitions of $H_2$ and the atomic fine-structure [S I] 25 μm, [Fe II] 26 μm, and [Si II] 35 μm lines (e.g., Neufeld et al. 2009; Dionatos et al. 2010). When hot water has been detected (Melnick et al. 2008), relatively few $H_2O$ transitions are seen, all in the LH module at λ > 29 μm. This water emission indicates low gas densities ~ $10^5$ cm$^{-3}$ and is consistent with non-dissociative shocks, in agreement with conclusions reached for water emission observed from outflows in the far-infrared (e.g., Giannini et al. 2001; Benedettini et al 2002; Nisini et al. 2010). A potential exception to the above is the possibility of a shock contribution to the OH emission seen in the T Tauri stars, given that similarly high rotational transitions of OH have been detected in one Herbig-Haro object (Tappe et al 2008; see §6.3). Dissociative shocks could explain some of the atomic lines observed in our sample. Emission from [Fe II] 26 μm, and sometimes [Si II] 35 μm, is present in stars with strong jets; the [Fe II] 17.9 μm transition is also strong in some of these stars, though this line is not apparent in outflows. In addition, the exceptionally strong [Ne II] emission from DG Tau shows a blueshift of 130 km s$^{-1}$ in our data and is likely to originate from a jet (see Guedel et al. 2010 on jet contributions to [Ne II] emission).

## 6.2 Water Emission

We find that the properties of the water emission are fairly similar among the sources in our sample, with a narrow range in the LTE temperature (with a mean of ~ 600 K) and $H_2O$ column densities ~ $10^{18}$ cm$^{-2}$ (see also Salyk et al. 2011). The uniformity in the water emission spectra (optically thick emission with similar excitation temperature) is remarkable. Given possible variations in the grain properties, UV and X-ray irradiation, accretion heating, transport processes of disk atmospheres, one might imagine that inner disk atmospheres and the water emission from them might be highly diverse. This does not appear to be the case.

The narrow temperature range suggests that $H_2O$ could play a role in regulating the temperature of the gaseous disk atmosphere. The large $H_2O$ line luminosities determined from the mid-infrared $H_2O$ spectra of disks demonstrate that water is a major gas coolant of the disk atmosphere (Pontoppidan et al. 2010a). At the same time, water can contribute to the gas heating through absorption of FUV radiation from the accretion shocks and near-infrared flux from the star. If absorption in the FUV bands of water dominates the gas heating, then water self-shielding will place an upper limit on the total column of warm $H_2O$, at $\sim 10^{18}$ cm$^{-2}$ (Bethell & Bergin 2009), a value that is similar to our derived water column densities.

The thermal-chemical disk atmosphere models of Glassgold et al. (2009) reproduce well the water temperatures and column densities that we derive. Particularly interesting is their finding that the molecular emitting layer has a temperature structure that does not change much over a range in disk radii (0.25-2AU), while maintaining water column densities of $10^{17}$ - $10^{18}$ cm$^{-2}$. Water heating and cooling were not included in their models; it would be interesting to add these processes to determine their effect on the temperature structure of the disk atmosphere.

### 6.3 OH Emission

Our detection of high-rotational OH transitions suggests the role of UV irradiation in producing the OH emission. The 14.65 micron feature, along with the other OH features in the SH spectrum, have upper energy levels several 1000 K above ground and relative intensities indicating a characteristic rotational temperature of ~4000 K (§ 4.2.2). Hot OH emission of this kind is prominent in the SH spectrum of the transition object TW Hya and has been suggested to result from the photo-dissociation of water by stellar UV (Najita et al. 2010). Photo-dissociation by FUV photons in the second absorption band of water ($\lambda$ = 1200-1400 Å) produces OH molecules in highly excited rotational levels but predominantly in the ground vibrational state (Harich et al. 2000).

The hot OH emission we measure in the SH module differs from the warm ($\leq 1000$ K) OH emission detected via lower rotational OH lines in the LH module (Carr & Najita 2008; Salyk et al. 2011). The presence of both hot and warm OH components can be understood if the OH molecules produced by photo-dissociation first emit hot (prompt) OH emission and then thermally relax to the temperature of the surrounding gas, producing cooler thermal emission (Najita et al. 2010).

Glassgold et al. (2009) suggested the possible role of UV irradiation in dissociating disk water vapor to produce abundant OH, noting that the OH column densities predicted by their thermal-chemical models fell far short of those reported for the warm (~500 K) OH emission for AA Tau (Carr & Najita 2008). Bethell & Bergin (2009) have shown that the column densities of the warm OH emission agree well with the predictions of a simple UV irradiated slab model in which OH is produced by the photo-dissociation of water in the disk surface layers. In their model, the maximum OH column is limited by the FUV opacity due to OH and $H_2O$ self-shielding.

Consistent with a role for UV radiation, we find that the 14.65 μm OH flux increases with stellar accretion rate (Figure 10), which is a measure of the UV flux of the system. We also found that the flux ratio of the 14.65 μm OH feature (non-thermal) and the 17 μm water feature (thermal) is relatively constant among a majority of sources in our sample (Figure 9), suggesting a close connection. One possibility is that they are both driven by stellar accretion, with the OH produced by UV dissociation and the water heated by UV radiation and/or accretion heating in the disk. We note that the highest accretion rate objects in our sample show OH emission but not $H_2O$ emission. This suggests that at sufficiently high FUV fluxes, $H_2O$ may be destroyed in favor of OH in the warm disk atmosphere (Bethell & Bergin 2009). Extending such an effect to the higher FUV fluxes for Herbig Ae/Be stars may explain the detection of OH and but not $H_2O$ emission from Herbig Ae/Be stars (Mandell et al. 2008; Pontoppidan et al. 2010a).

An alternative interpretation is that some fraction of the OH detected from the higher accretion rate objects in our sample may arise in shocks close to the star that are produced by the outflows in these systems (DG Tau, AS353A, V1331 Cyg). Tappe et al. (2008) detected highly excited OH emission in the *Spitzer* spectrum of young stellar outflow HH 211. That emission is comparable, but with higher rotational levels, to that observed here. They attributed the origin of the OH emission to $H_2O$ that is the photodesorbed from grains and then photodissociated by the UV radiation generated in the terminal outflow shock.

6.4 Organic Emission

The observed variation in the strength of the organic emission features (HCN, $C_2H_2$, $CO_2$) is difficult to interpret. The relative fluxes of the organic features and their strengths with respect to $H_2O$ vary, often by an order of magnitude or more (Fig. 9). The ratio of the derived column density of $C_2H_2$ to HCN shows a similar range (Table 4). A number of factors could potentially impact the chemistry or structure of the disk atmosphere and produce variations in the relative molecular fluxes. Some obvious physical processes include UV and X-ray irradiation of the disk, accretion heating, and grain settling. We examined the position of points in Figure 9 with respect to measured system parameters that track these processes to search for any patterns (e.g., whether all stars with strong $C_2H_2$ relative to $H_2O$ have high X-ray fluxes). We considered the mass accretion rate (Table 1) as a measure of UV radiation and accretion heating, the un-absorbed X-ray flux (Guedel et al. 2010), and the mid-infrared spectral index $n_{13-25}$ as a measure of grain settling (Furlan et al. 2006).

We did not find any obvious patterns that predict the behavior of the relative fluxes of the organic molecules with respect to these parameters. Perhaps larger datasets are required to reveal the influence of a particular parameter, particularly if multiple parameters play a role. Alternatively, the lack of an obvious predictor could indicate the importance of another parameter that has not been explored.

One possible factor, which lacks a direct observable, is variations in the C/O ratio of the gas. Such variations are expected to arise in the inner disk as a consequence of the growth and transport of icy material over sizes ranging from grains to planetesimals and protoplanets. As described by Ciesla & Cuzzi (2006 and references therein), when icy grains in the outer disk grow into approximately meter-sized bodies, they experience torques that lead them to migrate inward more rapidly than the surrounding gas. As they pass the "snow line" on their inward journey, they vaporize, pumping water (and oxygen nuclei) into the gas phase. If these bodies instead collide and grow to planetesimal size while they are still beyond the snow line, they drop out of the accretion flow and can thereby sequester water (and oxygen) in the outer disk.

In the former case, the hydration of the gaseous inner disk leads to a reduction in the C/O ratio. In the latter case, the sequestration of water in the outer disk leads to an enhancement in the C/O ratio. The detailed models of Ciesla & Cuzzi (2006) illustrate how the water abundance in the inner disk (and hence the C/O ratio) can be enhanced or reduced by an order of magnitude relative to its initial value, depending on the efficiency with which large bodies grow. Higher C/O ratios would tend to enhance the formation of organic molecules.

As the efficiency with which large bodies form may depend on the initial conditions (e.g., disk mass), we could expect a range in the C/O ratios of the inner disks among T Tauri stars of similar age, such as in our sample. Higher mass disks, which may have experienced more extensive planetesimal and protoplanet formation, might be expected to have a higher C/O ratio and enhanced abundances of organic molecules in the inner disk. Interestingly, the two sources in our sample with the largest $HCN/H_2O$ flux ratios (AA Tau and BP Tau) are the two sources with the largest disk masses (Figure 11 plots $HCN/H_2O$ vs. disk mass, using the disk masses in Table 1). Studies of larger samples are needed to explore the reality and implications of a relationship between $HCN/H_2O$ and disk mass.

## 7. ABUNDANCES

### 7.1 Inner Disk Abundances

Disk molecular abundances can provide valuable insights into disk chemical and physical processes. However, extracting abundances from the infrared molecular spectra is far from trivial. A simple approach is to compare the relative column densities of molecules derived from the LTE modeling, although this approach does have issues with regard to both interpretation and accuracy.

Firstly, the molecules measured in the *Spitzer* IRS spectra only provide a global average of the disk emission, and it is not known to what extent different molecules occupy the same volume, given the absence of spatial information. The LTE modeling implies comparable temperatures for $H_2O$, HCN and $C_2H_2$, suggesting that they are similar in their distributions, but the $CO_2$ emission may arise from regions of cooler gas. Eventually, line profiles obtained with high-resolution spectrographs can provide information on the relative radial distribution of the emission by using velocity and assumed Keplerian motion as a proxy for radius. However, various molecular species, even those arising from the same disk radii, could still have different vertical distributions in the atmosphere. Thermal-chemical models of disk atmospheres may provide insights into whether certain molecular species are likely to be co-spatial. The value of such insights depends critically on vetting the models by comparing their predicted structures with observations of disks. The data presented here provide a small step in that direction.

Secondly, it is worth noting that the infrared molecular emission is likely to probe only the atmosphere of the disk. The inner disks of classical T Tauri stars are optically thick at mid-infrared wavelengths, and the observed emission lines probably arise in a temperature inversion above the disk photosphere. A key question then is whether the composition in the line formation region is representative of the vertically integrated column of the disk. The chemistry in the upper disk atmosphere is likely to be different than that in deeper layers toward the mid-plane. Irradiation by UV and X-rays, in particular, is expected to be important; in addition, the densities are lower, and gas to dust ratios will be higher than in the midplane, if grain settling has occurred. Vertical chemical gradients could, however, be smoothed or reduced by mixing, if the timescale for vertical transport is short with respect to the key chemical timescales.

In contrast to the likely true physical situation, the model used to determine the column density and temperature from the spectra assumes a plane parallel layer for each molecule that is homogeneous in both vertical and radial directions and that the rotational and vibrational level populations are described by a single Boltzman distribution. In spite of its simplicity, the LTE model does a surprisingly reasonable job of fitting the observed spectra. There are differences between the calculated and observed $H_2O$ spectra, which suggest that some higher energy levels are sub-thermally populated (§ 4.2.2). This is not surprising given the real possibility that gas densities in the upper atmosphere could be too low to thermalize all the level populations. Radiative pumping may also contribute to the excitation, though there is no clear evidence that this is occurring (§ 4.2). Finally, it should be noted that the adopted fitting procedures and other modeling assumptions can also have considerable impact on the results (cf. Salyk et al. 2011).

With these known caveats, we examine the relative column densities of $H_2O$, HCN, and $C_2H_2$ for the T Tauri stars in our sample. In Figure 12 is plotted the column density ratio of $C_2H_2$ to HCN vs. the ratio of HCN to $H_2O$ for the five stars where we could model the HCN emission. The large error bars on the HCN/$H_2O$ ratio reflect the large parameter range allowed by the fits on the HCN column density. The lower end on the error bars corresponds to the optically thin solution with the assumption of equal HCN and $H_2O$ emitting areas. The best-fit points for the HCN column are a few percent that of $H_2O$, with an average ratio of 0.04. If we treat the upper and lower errors bars as the range of possible systematic effects in fitting the HCN band, then the likely range for HCN/$H_2O$ is 0.005 to 0.15. The smaller errors bars for the $C_2H_2$/HCN column density ratio result from the assumption that $C_2H_2$ and HCN have the same temperature and emitting areas. Under these assumptions, there is a real range in the $C_2H_2$ to HCN ratio, from 0.04 to 0.4, with an average of ~ 0.1.

## 7.2 Comparison to Envelopes, Outer Disks, Comets, and Hot Cores

We can compare the inner disk abundances described above with the abundances of gaseous protostellar envelopes, outer T Tauri disks, and comets. Since these systems probe the gas and solid phase abundances at larger radii than the inner disk, they represent chemical precursors that the inner disk ultimately inherits as a consequence of envelope infall, disk accretion, and solid-body migration. Such a comparison may lead to insights into the extent of chemical processing in disks.

The HCN/CO abundance ratios for inner disks appear to be a few orders of magnitude larger than the values for low-mass protostellar envelopes and the outer disks of T Tauri stars. Prestellar cores, Class 0 sources, and Class 1 sources, which are the evolutionary precursors of T Tauri stars, typically have HCN/CO mass ratios of $10^{-4}$ or smaller in the gas phase, as measured in millimeter emission lines (Jørgensen et al. 2004). The outer disks surrounding T Tauri stars are also found to have HCN/CO column density ratios of a few times $10^{-4}$ or smaller (Thi et al. 2004; Dutrey et al. 1997). In comparison, our inferred HCN abundance relative to $H_2O$ for inner disks is approximately a few percent. If we assume that the disks in our sample have CO/$H_2O$ column density ratios ~1, as found for AA Tau and other T Tauri stars (Carr & Najita 2008; Salyk et al. 2011), then the HCN/CO ratios for inner disks are also a few percent. Similarly high HCN/CO ratios are derived from infrared absorption measurements towards two low-mass young stars where the absorption is interpreted as arising in the inner region of a disk seen close to edge-on: HCN/CO is 0.025 in IRS 46 (Lahuis et al. 2006a) and 0.014 in GV Tau N (Gibb et al. 2007, 2008). Some of the difference between the gas phase HCN/CO ratio of inner disks and those of envelopes and outer disks may be accounted for by freeze out onto grains, which is thought to be substantial (e.g., Jørgensen et al. 2004).

Cometary abundances provide insight into what was frozen out onto grains, because comets are made up of icy material that was originally present in the Jupiter-Saturn region and in the outer solar nebula (the Kuiper Belt) during the formation of the solar system. Analyses of near-infrared spectroscopy of comets find a mean HCN/$H_2O$ abundance ratio of 0.0026 and a mean $C_2H_2$/$H_2O$ abundance ratio of 0.0024; thus the $C_2H_2$/HCN ratio is typically ~ 1 (DiSanti et al. 2009; DiSanti & Mumma 2008). In comparison, the HCN/$H_2O$ ratios we find for inner disks are all systematically higher than those found for comets, by approximately an order of magnitude on average (see Figure 12). The $C_2H_2$/HCN abundance ratios for inner disks are also systematically lower than those of comets by a factor of 2–20, or approximately an order of magnitude on average.

A larger HCN/$H_2O$ ratio for gaseous inner disks, compared to both comets and gaseous outer disks, would suggest that the abundances of inner disks are not simply the result of gas and icy material migrating from the outer to the inner disk. Instead, further gas-phase processing may occur within the snow line following the evaporation of icy material.

Our results can also be compared to the abundances of hot cores, the molecular envelopes surrounding high-mass protostars. Hot cores have long been known to show a rich chemistry with complex organic molecules. The same molecules we observe in emission in the T Tauri stars have also been detected in absorption in mid-infrared bands towards several hot cores using *ISO* (Lahuis & van Dishoeck 2000; Boonman & van Dishoeck 2003). The derived gas temperatures for the absorbing gas are in the same range as we found for the T Tauri stars. The ratios of reported absorption columns are plotted in Figure 12. Typical values are HCN/$H_2O$ ~ 0.01, larger than those of comets and within the range found here for inner disks, and $C_2H_2$/HCN ~ 0.5 (at the upper end of the range inferred here for inner disks).

Thus the *average* inner disk atmosphere appears to have similar or slightly higher HCN abundance with respect to $H_2O$ than do hot cores. However, the HCN abundance would be at least an order of magnitude larger if CO were taken as the reference molecule. This is because the infrared absorption columns towards hot cores have an average CO/$H_2O$ column density ratio of 10, whereas the disks in our sample are more likely to have CO/$H_2O$ column density ratios ~1, (Carr & Najita 2008; Salyk et al. 2011).

Similar chemical processes could be in operation in hot cores and inner disks. In hot cores, the warm gas phase follows the sublimation of molecules from icy grain mantles, where the rich chemistry that is observed is the result of some combination of grain-surface and warm gas-phase reactions. Inner disks present an analogous situation, in that the warm gas-phase molecules we observe represent material that was desorbed from grains as it migrated inward from the outer disk. Because of these similarities, hot core chemistry may serve as a rough guide to the kind of chemical synthesis that may be occurring in inner disks. Numerous complex molecules are detected in hot cores (e.g., Herbst & van Dishoeck 2009). Similar complex species could be present in inner disks.

## 7.3 Comparison to chemical models of inner disks

Several features of the molecular emission we observe are reproduced in recent chemical models of inner disks. The thermal-chemical calculation of Glassgold et al. (2009) is able to account for the large column densities ($\sim 10^{18}$ cm$^{-2}$) and warm temperatures ($\sim 500$ K) of the water emission by invoking $H_2$ formation on warm grains and (accretion-related) mechanical heating in the disk atmosphere. Grain settling also plays a role in achieving large column densities of warm molecular gas, by reducing the gas-grain cooling in the atmosphere (Glassgold et al. 2004, 2009).

In the thermal-chemical protoplanetary disk model of Woitke et al. (2009; PRODIMO), water is also present in the inner disk atmosphere although at much lower column densities ($<10^{14}$ cm$^{-2}$ above 200 K at 1 AU) than we measure. Willacy & Woods (2009) also calculate the gas temperature in their model of disk atmospheres. They find very large columns of warm (>1000 K) $H_2O$ ($\sim 10^{19}$ cm$^{-2}$) at 1 AU, much larger than those measured here. The different dominant sources of heating and assumptions regarding grain abundances in these models may be partly responsible for the very different predictions.

The disk chemistry calculation of Agúndez et al. (2008) considers the formation of organic molecules as well as water. In lieu of a thermal calculation, they adopt an existing model for the temperature and density structure of the disk (D'Alessio et al. 1998, 1999), which is then irradiated by a (primarily) interstellar FUV field. At 1 AU, the resulting HCN vertical column density at the disk surface is $\sim 2 \times 10^{16}$ cm$^{-2}$, with a $C_2H_2$ column density that is $\sim 10$ times smaller and $H_2O$ and CO columns $\sim 2 \times 10^{17}$ cm$^{-2}$ and $\sim 8 \times 10^{17}$ cm$^{-2}$. These values are similar to the values we measure for inner disk atmospheres. At 2 AU, the HCN vertical column density at the disk surface is much smaller, $\sim 5 \times 10^{13}$ cm$^{-2}$, with a $C_2H_2$ column density that is $\sim 5$ times smaller. If these results are a guide to what to expect in disk atmospheres, the declining HCN and $C_2H_2$ abundances beyond 1 AU suggest that HCN and $C_2H_2$ emission would be limited to the inner 1-2 AU of the disk, consistent with our results.

An important difference between the observations and the Agúndez et al. disk atmosphere model is the gas temperature. Because the gas temperature in the calculation is taken to equal the dust temperature, the surface molecular layers are much cooler (≤ 300 K) than are observed (~ 500–700 K). Warmer disk atmospheres would tend to further enhance the molecular abundances (Fig. 1 of Agúndez et al.).

Willacy & Woods (2009) also study HCN and $C_2H_2$. At 1 AU, they find vertical columns of warm HCN (~$4 \times 10^{16}$ cm$^{-2}$) that are comparable to those measured here, accompanied by much smaller columns of $C_2H_2$ (~$9 \times 10^{8}$ cm$^{-2}$) than we measure. Both species are present at warmer temperatures (>1000 K) than we measure. The differences between these models and the observations may help to guide the future development of disk atmosphere models.

## 8. SUMMARY

One notable discovery of the *Spitzer* mission is that molecular emission is common in the mid-infrared spectra of T Tauri stars. In the sample of 11 classical T Tauri stars studied in this report, we find high detection rates of $H_2O$ and OH rotational lines and ro-vibrational bands of $CO_2$, HCN, and $C_2H_2$. *Spitzer* spectroscopy of larger samples of T Tauri stars (Pontoppidan et al. 2010a) produce similar results. Our sample shows significant star-to-star variation (up to an order of magnitude) in the absolute molecular fluxes, the equivalent widths, and the relative flux ratios of different molecules.

We estimated gas temperatures, column densities and emitting areas using an LTE slab model. For $H_2O$, we find a similar temperature for each object, about 600 K, a narrow range for line-of-sight column density, ~ $10^{18}$ cm$^{-2}$, and projected emitting areas of radius ~ 1 AU. Comparable or potentially higher temperatures are derived for HCN, in the range of 500-800 K. The $C_2H_2$ Q-branch is consistent with having the same temperature as HCN. The best fits to the HCN spectra require a smaller emitting area for HCN than for $H_2O$, though similar emitting areas are not ruled out. Fits to the $CO_2$ Q-branch are more uncertain, but imply lower temperatures than the other molecules (100-600 K), with an average best-fit temperature of 350

K. This suggests that the $CO_2$ emission arises in a cooler volume in the disk, either at larger radii or at a different vertical height.

The gas temperatures, emitting areas, and high critical densities for the observed $H_2O$ transitions are consistent with gas in a disk atmosphere at radii within 2 AU of the star. Hence, the observed mid-infrared molecular emission traces gas in the inner planet formation region, radii that correspond to the terrestrial planet region in the Solar System and that fall within the generally accepted location of the snowline in protoplanetary disks.

We discuss how the $HCN/H_2O$ or $HCN/CO$ column density ratios are roughly a few percent, which is about an order of magnitude larger than the average ratio derived for comets and at least two orders of magnitude larger than those measured for the outer disks of T Tauri stars and low-mass protostellar cores. The $HCN/H_2O$ ratio for inner disks is comparable to the ratios derived for hot molecular cores around massive protostars, but the $HCN/CO$ ratios could be an order of magnitude larger. These comparisons suggest a picture where the gas-phase abundances in the inner disk are the result of chemically active disks, rather than simple infall of material from the protostellar envelope and inward migration through the outer disk.

The $H_2O$ emission spectra are remarkably similar from star to star, which is reflected in the narrow range of model parameters and implies limited variation in $H_2O$ excitation conditions and optical depth. High-rotational transitions of OH are present in all spectra and are likely due to prompt emission of OH following FUV photo-dissociation of water. The role of UV radiation and a connection between OH and $H_2O$ are supported by the increase in OH emission flux with increasing stellar accretion rate and the relatively small scatter in the $OH/H_2O$ flux ratio. We propose that some of the wide variation in the relative emission strengths of the organic molecules with respect to $H_2O$ could be related to star-to-star differences in the C/O ratio (the oxygen fugacity) of the inner disk gas that affects the organic chemistry. Differing C/O ratios could result from enhancements or depletions of water in the inner disk depending on the efficiency of the growth of large bodies beyond the snowline (Ciesla & Cuzzi 2006). The objects in our sample with the largest $HCN/H_2O$ flux ratios also have the largest disk masses. We speculate that this can be understood if higher mass disks are more efficient at forming planetesimals and sequestering water in the outer disk, leading to higher C/O ratios and enhanced abundances of organic molecules in the inner disk.

This work is based on observations made with the Spitzer Space Telescope, which is operated by the Jet Propulsion Laboratory, California Institute of Technology under a contract with NASA. Support for this work was provided by NASA. Basic research in infrared astrophysics at the Naval Research Laboratory is supported by 6.1 base funding.

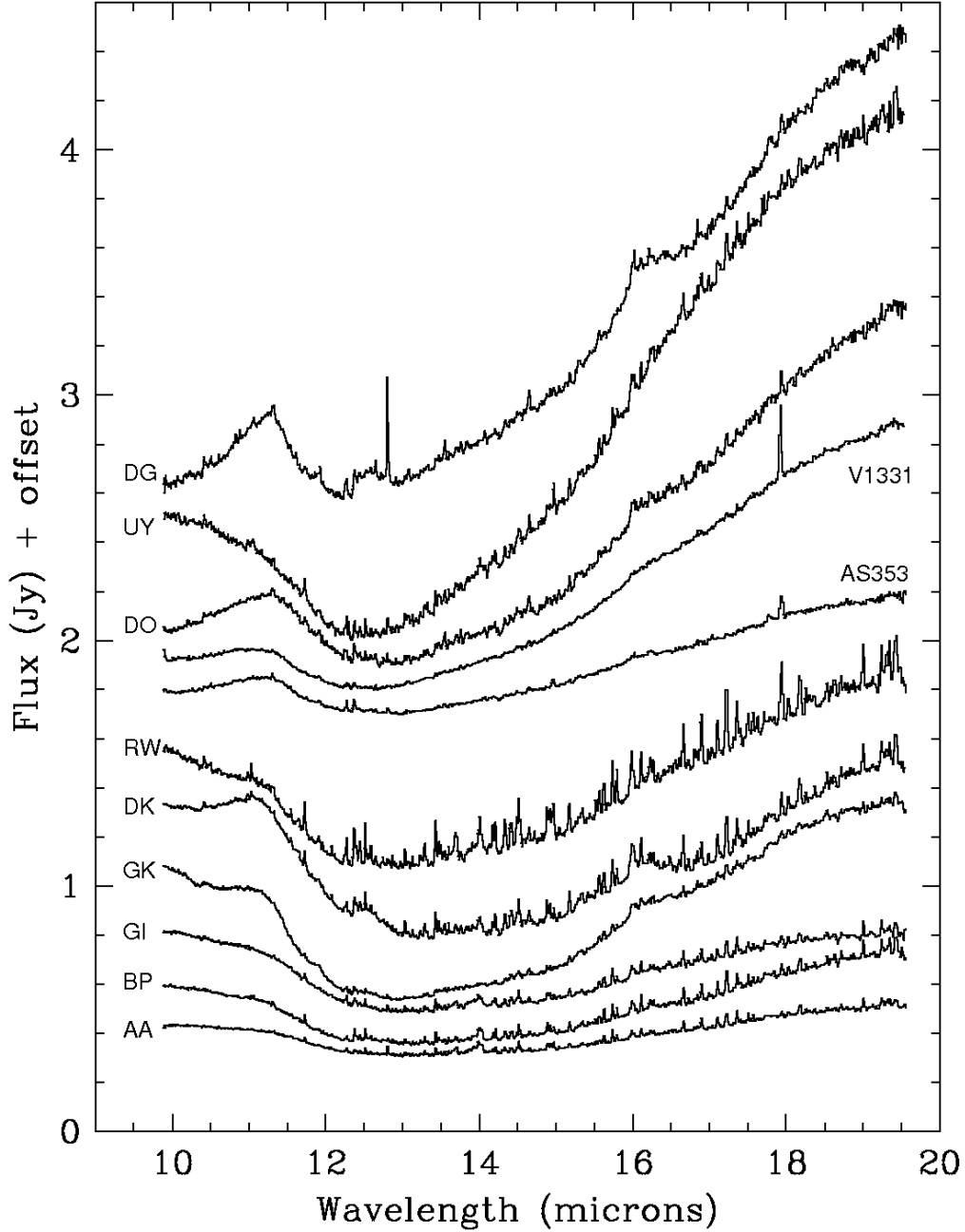

Fig. 1.— *Spitzer* short-high (SH) IRS spectra for the entire sample. For the purpose of plotting, offsets have been applied to the spectra of several stars. The offsets in Jy are: -0.25 for GI Tau, -0.20 for GK Tau, +0.05 for DK Tau, -0.35 for RW Aur, -0.40 for UY Aur, +0.85 for V1331 Cyg, and +0.60 for AS 353A. For DG Tau, the spectrum was also divided by a linear function in order to reduce the spectral slope and allow plotting on the same figure; hence, the relative line fluxes and the spectral slope across the SH band are reduced by nearly a factor of two from the true spectrum for DG Tau.

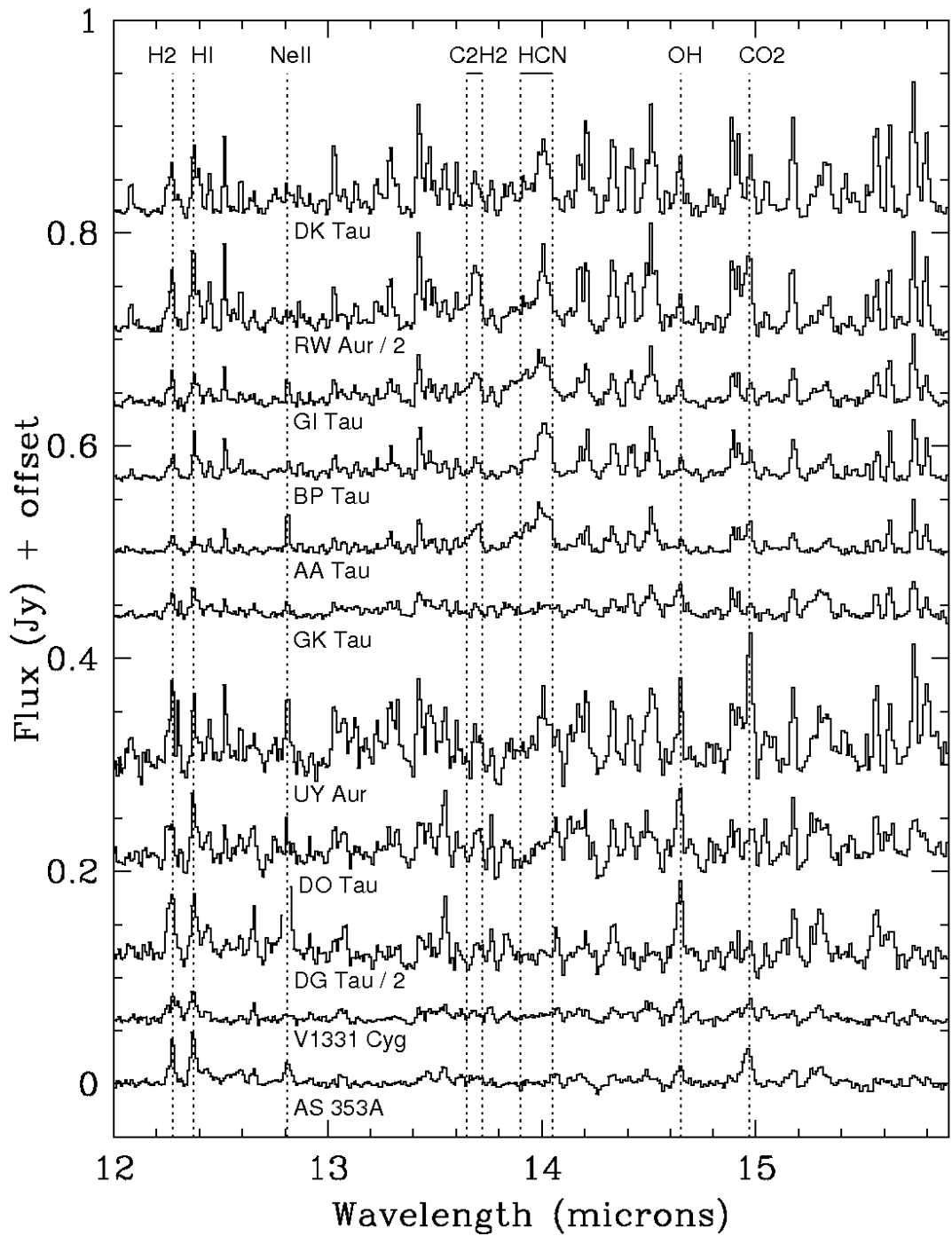

Fig. 2.— The SH spectra in the 12–16 μm region after subtraction of the continuum. The spectra for RW Aur and DG Tau have been divided by 2. The wavelength positions are indicated for the Q-branches of HCN, $C_2H_2$, and $CO_2$, the S(2) rotational transition of $H_2$, the 12.37 μm line of H I, the 12.81 μm [Ne II] line, and the 14.64 μm rotational feature of OH. The remaining emission features are mostly due to rotational transitions of $H_2O$ with some additional OH features.

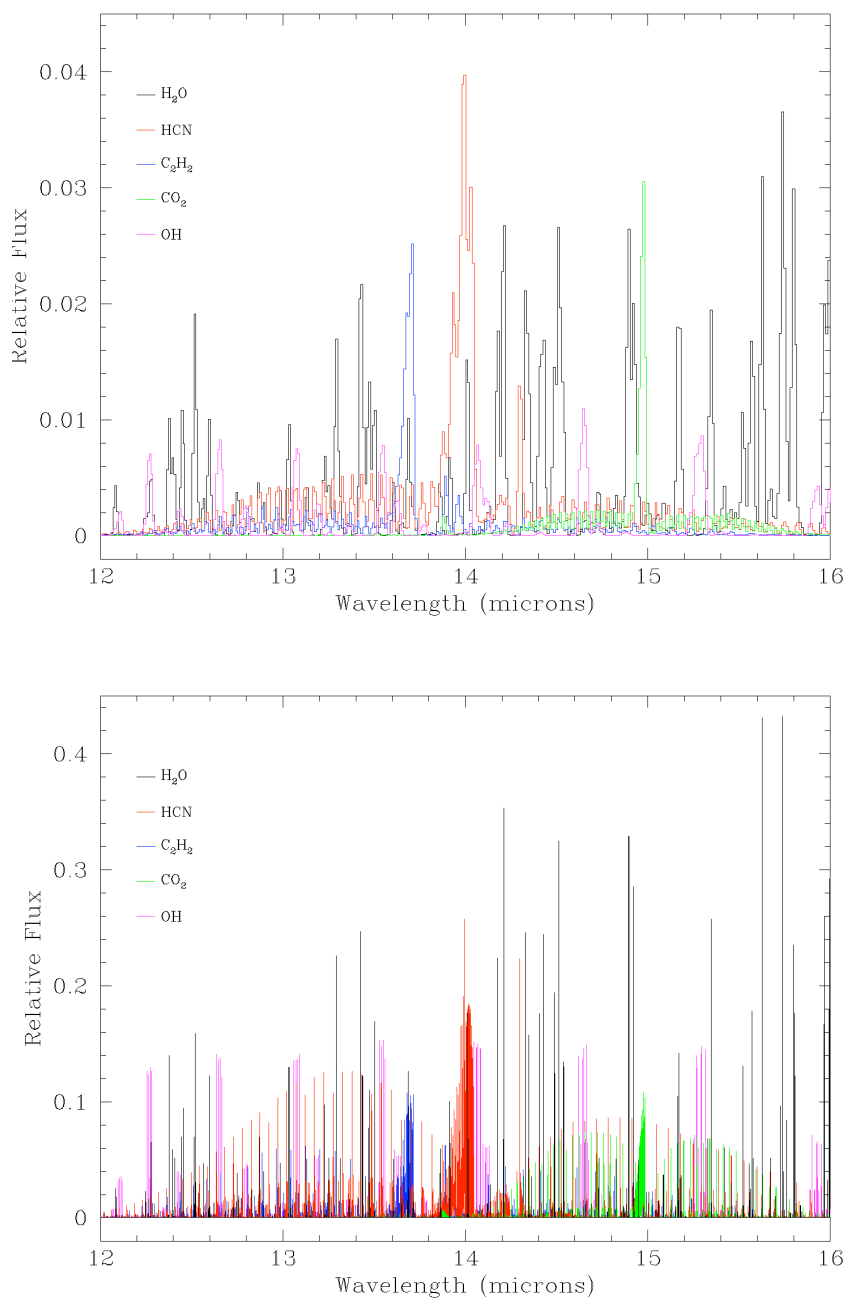

Fig. 3.— Synthetic spectrum for the 12–16 μm region, showing the individual contributions from each of the detected molecules. The temperatures, column densities and areas are characteristic of those derived for T Tauri stars in this paper. The $H_2O$ spectrum is plotted as a black histogram, HCN as red, $C_2H_2$ as blue, $CO_2$ as green, and OH as magenta. (a) For the resolution and sampling of the IRS SH module. (b) For a resolving power of R=30,000. The $H_2O$ flux was divided by 3 in order to display it adequately with respect to the other molecules.

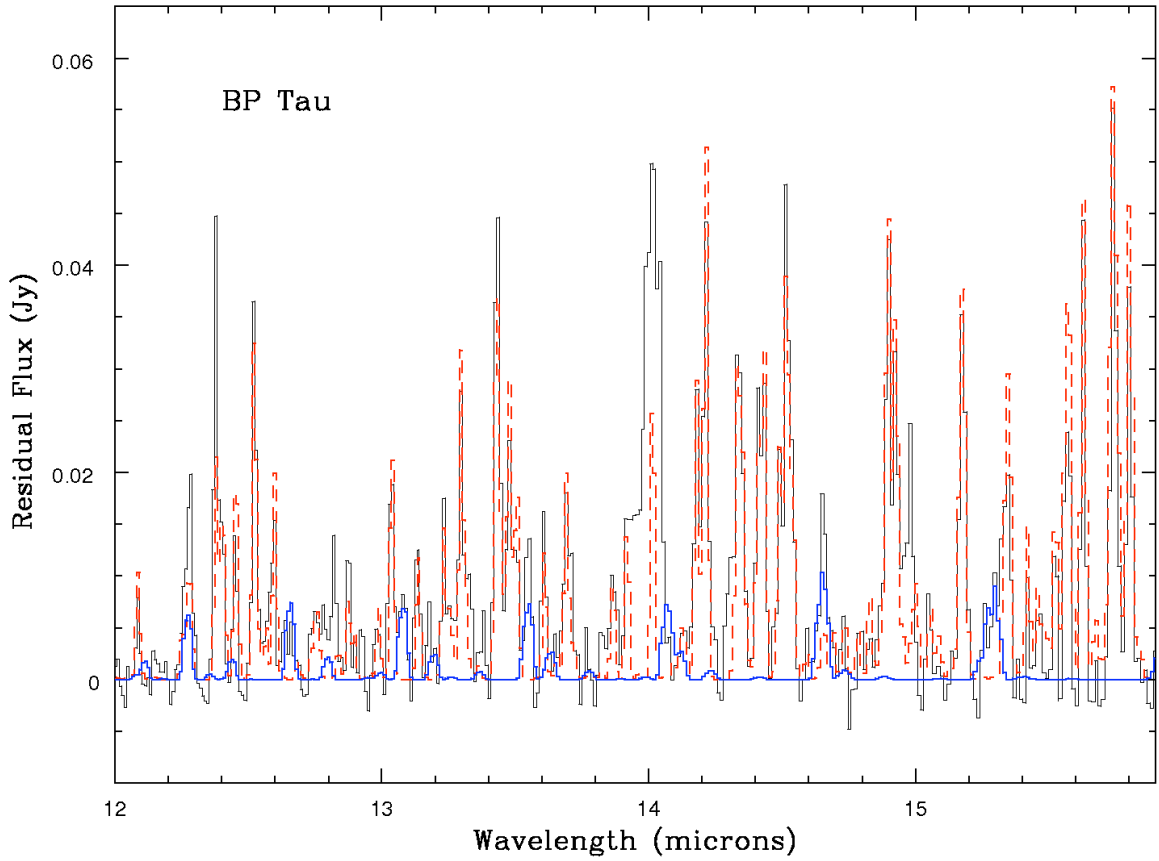

Fig. 4.— The synthetic $H_2O$ model spectrum for BP Tau (dashed red line) over-plotted on the observed continuum subtracted spectrum. The model parameters are given in Table 2. The solid bold (blue) line shows the model OH spectrum used for BP Tau.

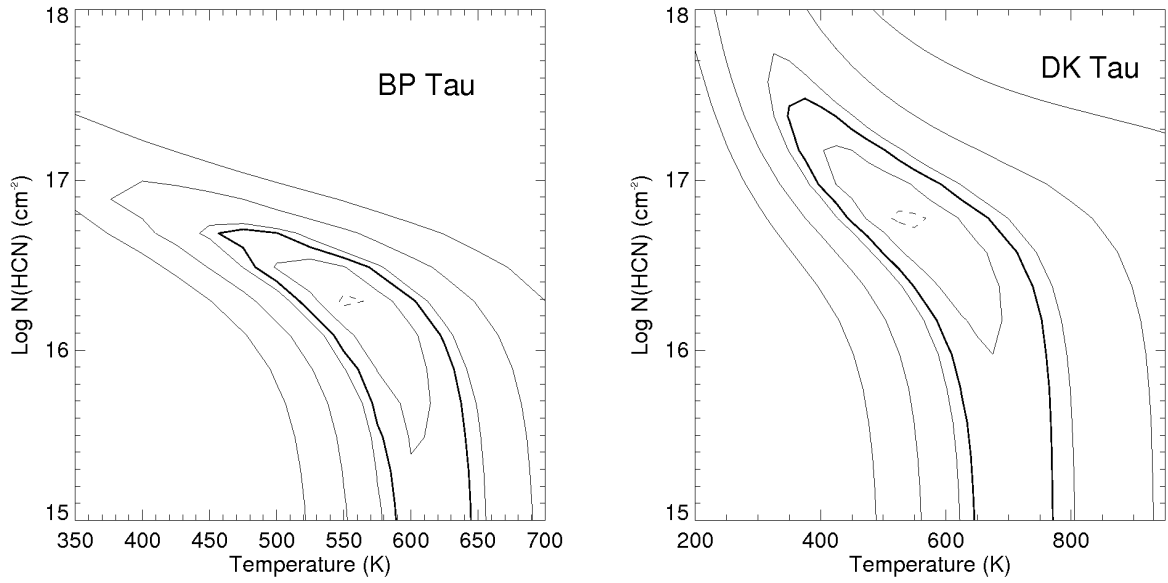

Fig. 5.— Contour plots of $\chi^2$ as a function of N(HCN) and T for fits to the HCN Q-branch in BP Tau and DK Tau. The bold line is the 90 % confidence contour, the dashed line indicates the location of minimum $\chi^2$, and the thin solid lines correspond to 1, 2, 3, and 4 $\sigma$.

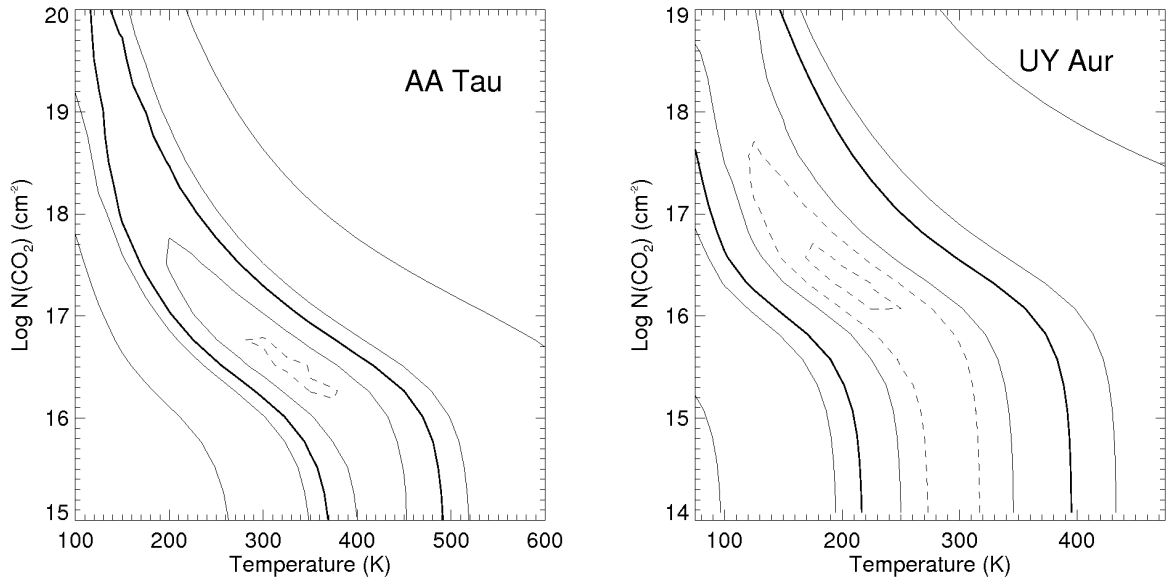

Fig. 6.— Contour plots of $\chi^2$ as a function of $N(CO_2)$ and T for fits to the' $CO_2$ Q-branch in AA Tau and UY Aur. The bold line is the 90 % confidence contour, the dashed line indicates the location of minimum $\chi^2$, and the thin solid lines correspond to 1, 2, and 3 $\sigma$.

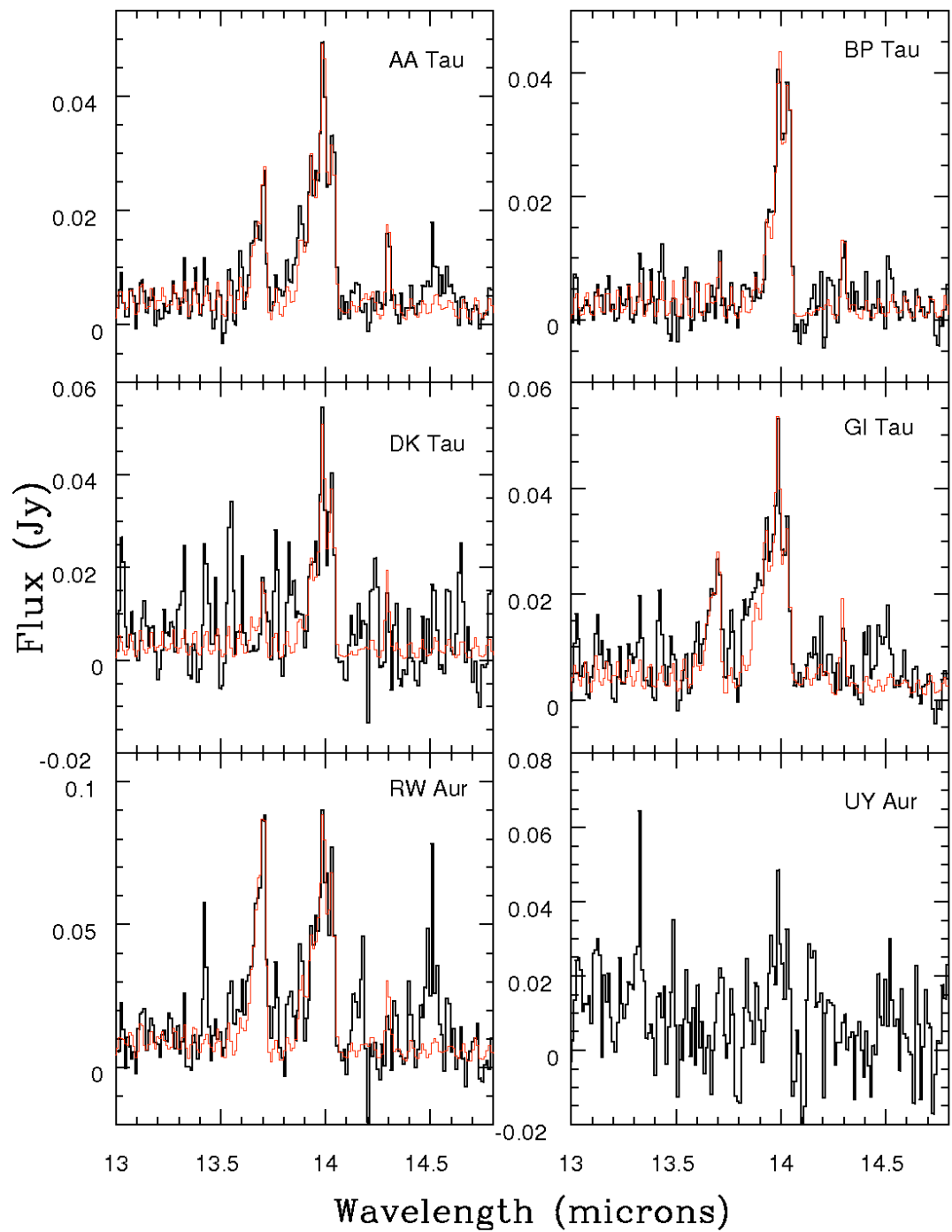

Fig. 7.— Synthetic spectra for HCN and $C_2H_2$ over-plotted on the data. The data (thick black histogram) are the continuum-subtracted spectra (Fig. 1) with the best $H_2O$ + OH model subtracted. The model spectra (thin red histogram) are the best-fit parameters for HCN and $C_2H_2$ in Tables 3 and 4. The residual spectrum for UY Aur is plotted here to show the marginal detection of HCN, but a model spectrum was not calculated.

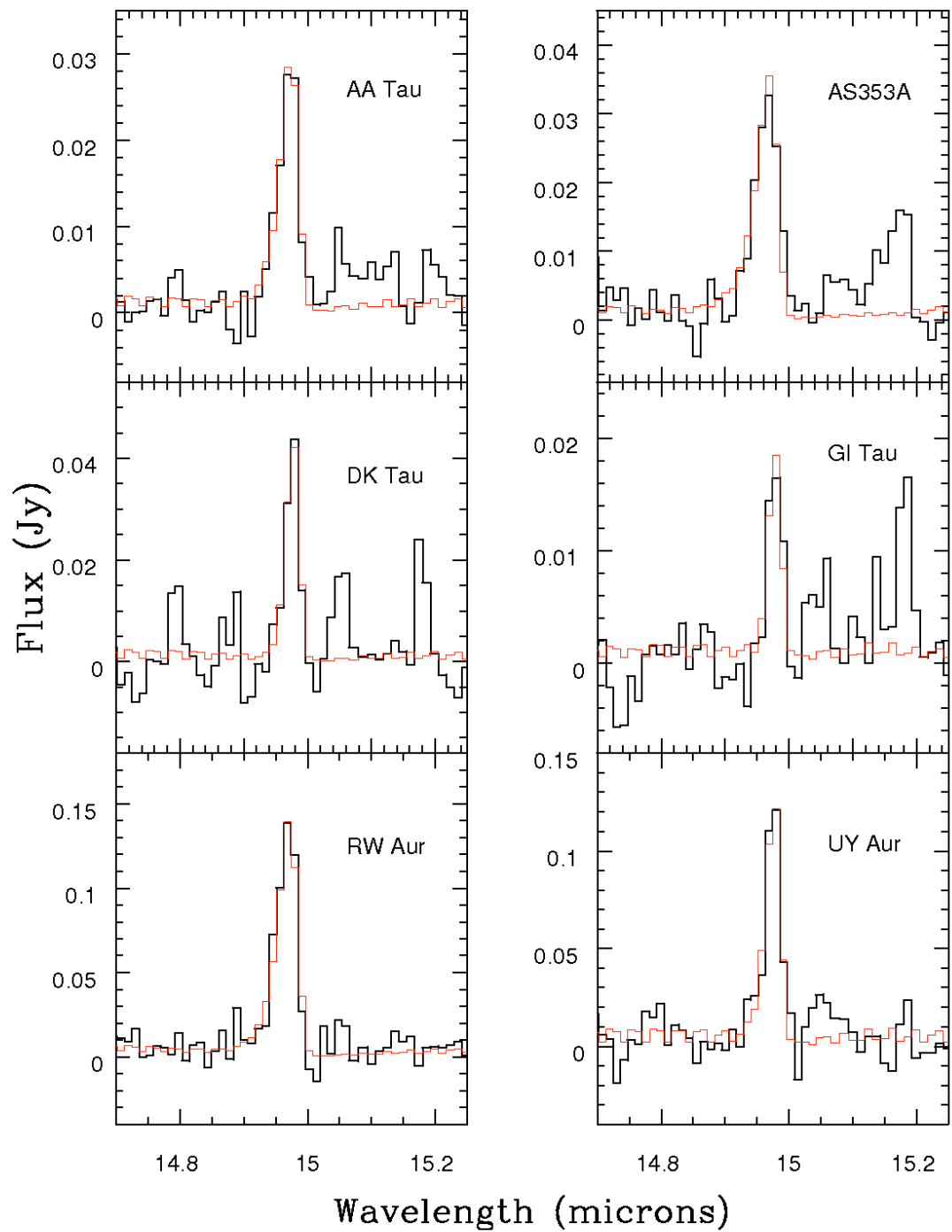

Fig. 8.— Synthetic spectra for $CO_2$ over-plotted on the data. The data (thick black histogram) are the continuum-subtracted spectra (Fig. 1) with the best $H_2O$+OH+HCN+$C_2H_2$ model subtracted. The model spectra (thin red histogram) use the minimum $\chi^2$ temperature solution in Table 5.

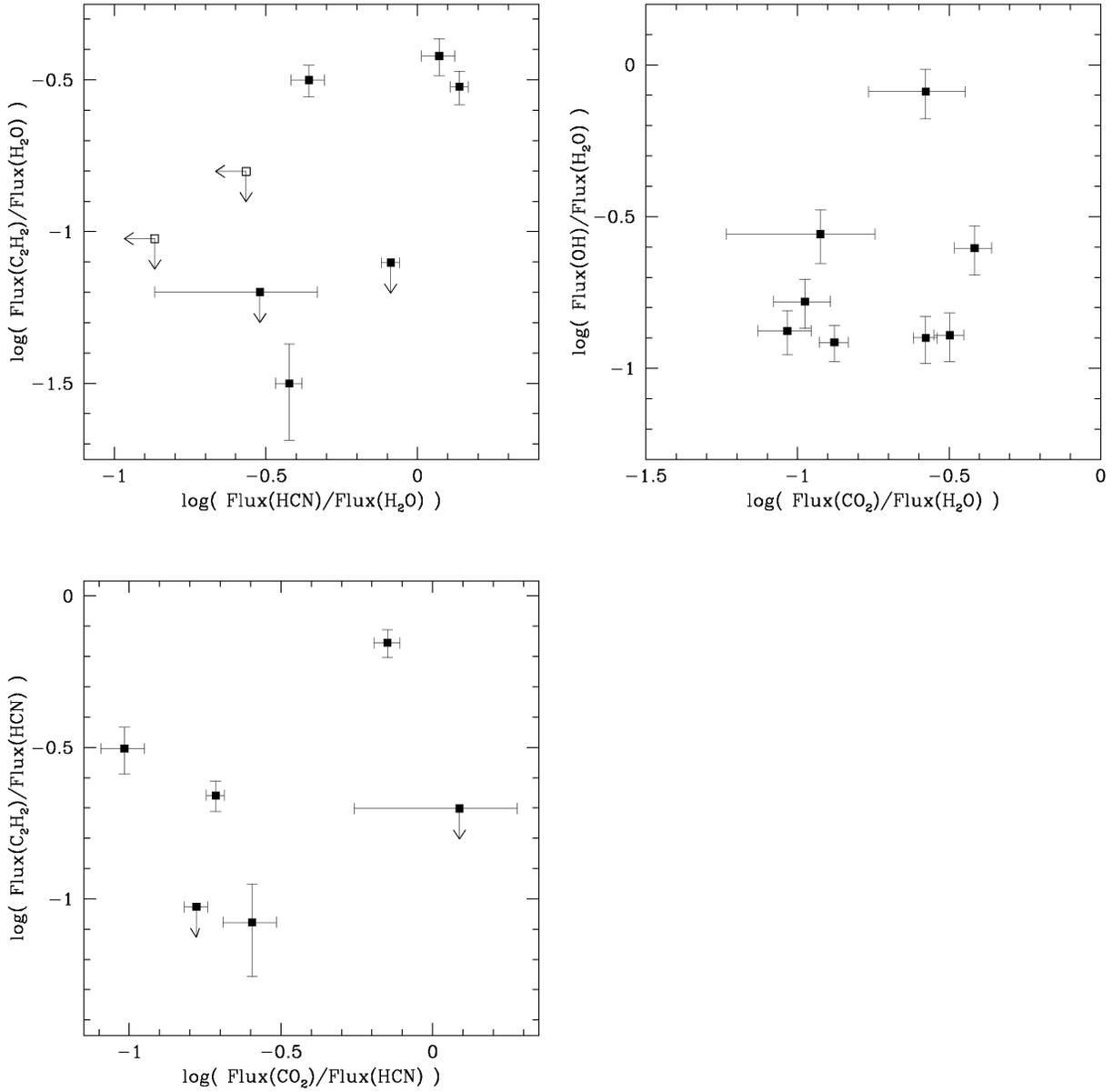

Fig. 9.— Log-log plots of different flux ratios against one another for measured molecules. All limits (arrows) are 2-σ. (a) Flux ratio of $C_2H_2$/$H_2O$ vs. HCN/$H_2O$. (b) Flux ratio of OH/$H_2O$ vs. $CO_2$/$H_2O$. (c) Flux ratio of $C_2H_2$/HCN vs. $CO_2$/HCN.

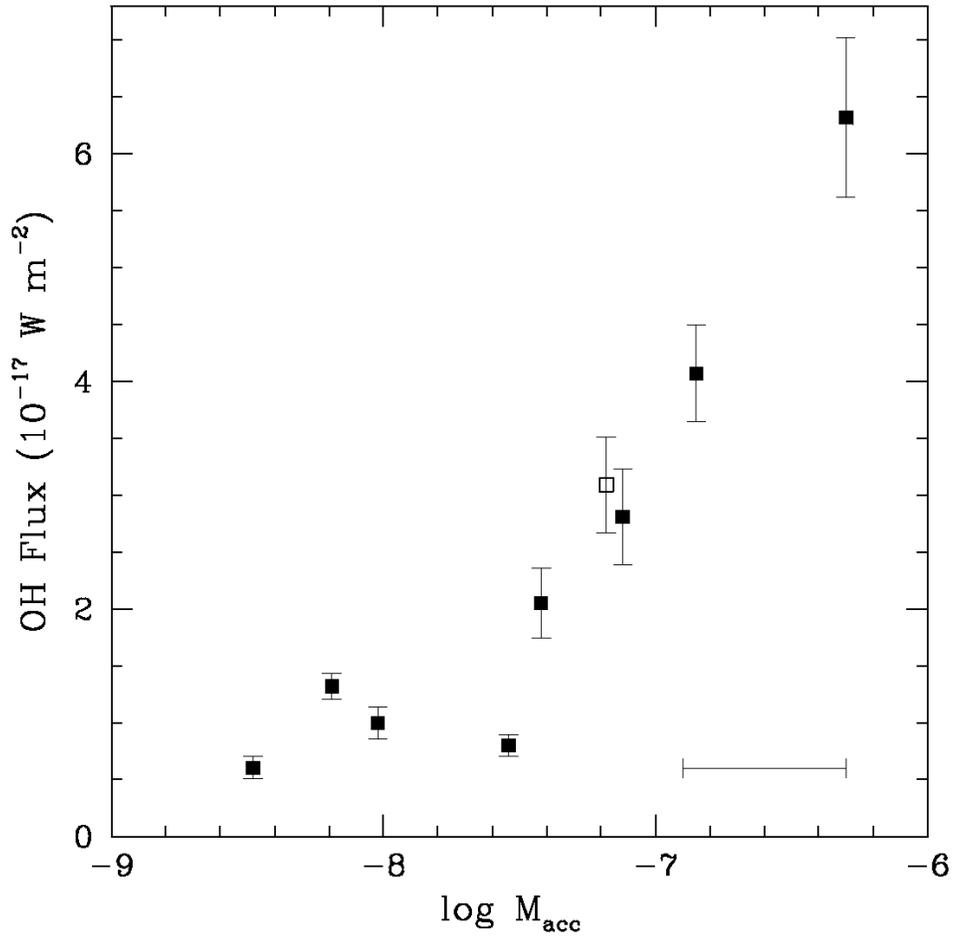

Fig. 10.— The OH emission flux vs. log of accretion rate. The accretion rate is in $M_{sun}$ yr$^{-1}$. The open square is for the binary UY Aur. The error bar in the lower right suggests a possible range in variability in the accretion rate.

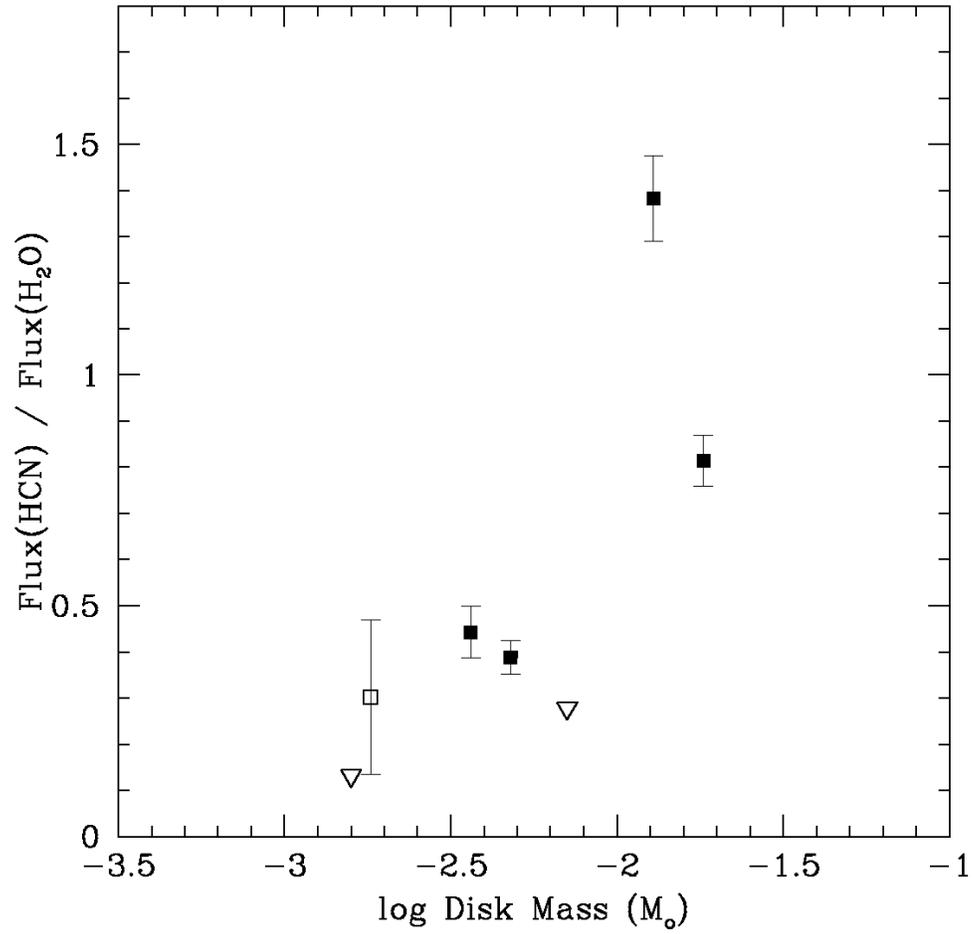

Fig. 11.— The flux ratio of HCN to H₂O plotted against disk mass. Inverted triangles are upper limits (2-σ) and correspond to stars with non-detections of HCN emission. The open square is for the binary UY Aur, which also has a marginal detection for HCN.

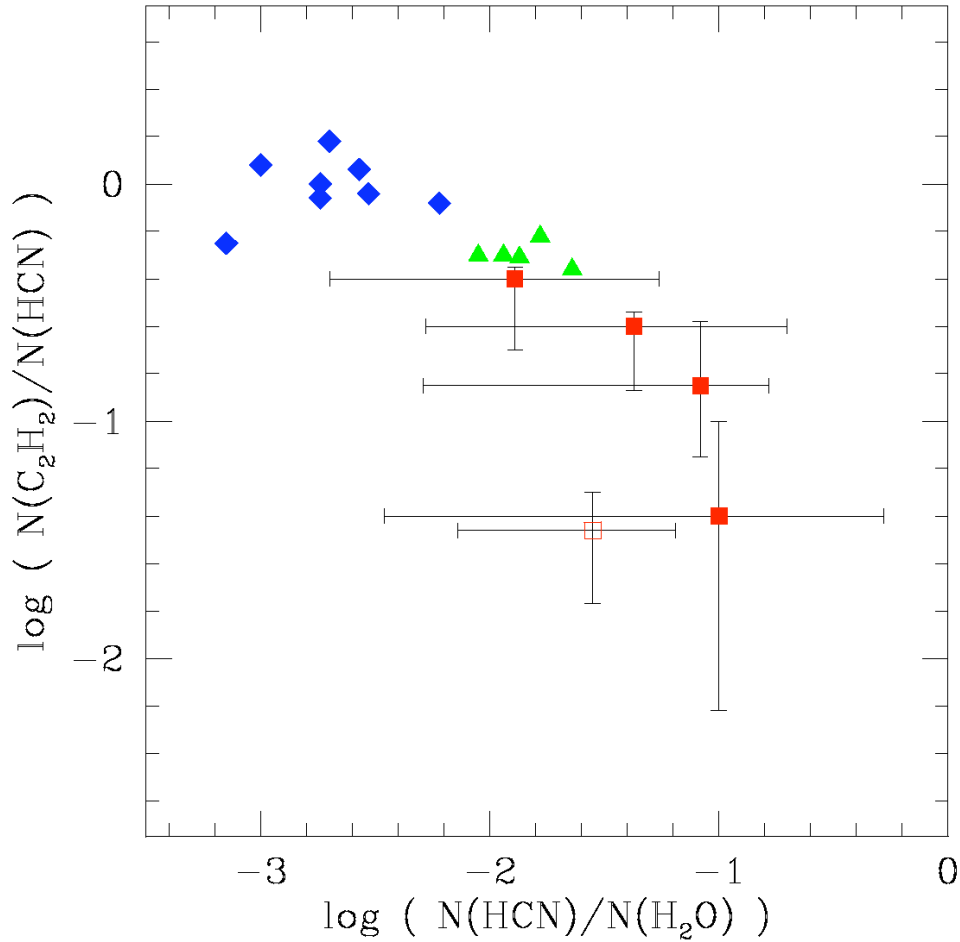

Fig. 12.— Line-of-sight column density of $C_2H_2$ to HCN plotted against the column density of HCN to $H_2O$. The squares are the column density ratios for T Tauri star disks from the LTE slab models in this paper. BP Tau, plotted as an open square, has an upper limit on the $C_2H_2$/HCN ratio. The lower error-bar on HCN/$H_2O$ for each star corresponds to the optically thin solution with equal emitting areas for HCN and $H_2O$. The triangles are column density ratios for hot molecular cores around massive protostars derived from mid-infrared absorption measurements. The diamonds are ratios of the abundances in comets as determined from near-infrared spectroscopy. Data for hot cores and comets are from references given in text.

**Table 1**

**Observations and Sample Parameters**

| Object | FrameTime[a] (sec) | Cycles[b] | Flux[c] (Jy) | SpT | log $M_{acc}$ ($M_{sun}$ yr$^{-1}$) | Disk Mass (log $M_{sun}$) |
|---|---|---|---|---|---|---|
| AATau | 31.5 | 24 | 0.32 | K7 | -8.5 | -1.89 |
| BPTau | 31.5 | 24 | 0.37 | K7 | -7.5 | -1.74 |
| DGTau | 6.3 | 24 | 5.25 | K7 | -6.3 | -1.62 |
| DKTau[d] | 31.5 | 16 | 0.75 | K7 | -7.4 | -2.32 |
| DOTau | 31.5 | 16 | 2.00 | M0 | -6.8 | -2.15 |
| GITau | 31.5 | 8 | 0.75 | K6 | -8.0 | … |
| GKTau | 31.5 | 8 | 0.79 | K7 | -8.2 | -2.80 |
| RWAur[d] | 6.3 | 26 | 1.48 | K3 | -7.1 | -2.44 |
| UYAur[e] | 6.3 | 20 | 2.66 | K7 | -7.2 | -2.74 |
| AS353A | 31.5 | 16 | 1.15 | K5 | -6.3 | … |
| V1331Cyg | 31.5 | 16 | 1.06 | C | -6.1 | … |

**Notes.** [a] Total SH integration time on source = 2x(Frame Time)xCycles.

[b] Number of on-source cycles. Number of cycles in off position is 1/2 that on-source, except for GI Tau and GK Tau, where the on and off source cycles are equal.

[c] Continuum flux at 14 μm measured in this work.

[d] Unresolved binary within the IRS slit, with secondary contributing < 10% of the broadband mid-infrared flux.

[e] UY Aur is an unresolved binary within the IRS slit with both components contributing similar amounts to the broadband mid-infrared flux.

**Table 2**

**Results for H₂O Modeling**

| Object | T (K) | N(H$_2$O)[a] ($10^{16}$ cm$^{-2}$) | R$_e$[b] (AU) |
|---|---|---|---|
| AA Tau | 575 (50) | 78 (20) | 0.85 (0.12) |
| BP Tau | 650 (50) | 78 (20) | 0.83 (0.09) |
| DK Tau | 650 (50) | 60 (12) | 1.27 (0.15) |
| GI Tau | 575 (50) | 42 (12) | 1.16 (0.20) |
| RW Aur | 600 (80) | 155 (37) | 1.49 (0.24) |
| UY Aur | 600 (100) | 179 (119) | 1.04 (0.45) |

[a] Line-of-sight column density

[b] Radius of projected emitting area

**Table 3**

**Results for HCN Modeling**

| Object | Best Fit[a] | | | Optically thick[b] | | | Optically Thin[c] | |
|---|---|---|---|---|---|---|---|---|
| | T (K) | N(HCN) ($10^{16}$ cm$^{-2}$) | $R_e$ (AU) | T (K) | N(HCN) ($10^{16}$ cm$^{-2}$) | $R_e$ (AU) | T (K) | N(HCN) ($10^{16}$ cm$^{-2}$) |
| AA Tau | 690 | 6.5 | 0.28 | 570 | 13.0 | 0.28 | 840 | 0.41 |
| BP Tau | 550 | 2.2 | 0.50 | 490 | 5.1 | 0.47 | 605 | 0.56 |
| DK Tau | 540 | 6.0 | 0.37 | 350 | 31.0 | 0.54 | 705 | 0.20 |
| GI Tau | 850 | 1.8 | 0.42 | 680 | 8.4 | 0.27 | 890 | 0.22 |
| RW Aur | 690 | 2.1 | 0.59 | 500 | 8.6 | 0.51 | 745 | 0.27 |

**Notes.** Column densities are line-of-sight, and radii are that of the projected emitting area.

[a] Minimum chi-square solution.

[b] Lower bound on T and upper bound on N(HCN).

[c] Optically thin solution with $R_e$ = R(H$_2$O) from Table 2.

**Table 4**

**N(C$_2$H$_2$) / N(HCN) Ratio**

| Object | Best Fit | Optically thick | Optically Thin |
|--------|----------|-----------------|----------------|
| AA Tau | 0.14     | 0.07            | 0.26           |
| BP Tau | <0.035   | <0.017          | <0.051         |
| DK Tau | 0.04     | 0.006           | 0.10           |
| GI Tau | 0.25     | 0.14            | 0.29           |
| RW Aur | 0.38     | 0.19            | 0.43           |

**Notes**. Ratio of C$_2$H$_2$ to HCN column density for the same cases in Table 3.

**Table 5**

**Results for $CO_2$ Modeling**

| Object | $T^a$ (K) | $T_{range}^b$ (K) | $R_{min}^c$ (AU) |
|---|---|---|---|
| AA Tau | 325 | 120-490 | 0.8 |
| BP Tau | 425 | 75-800 | 0.1 |
| DK Tau | 220 | 75-300 | 2.0 |
| GI Tau | 100 | 75-360 | 0.9 |
| RW Aur | 510 | 140-680 | 0.6 |
| UY Aur | 200 | 75-390 | 1.5 |
| AS353A | 600 | 160-800 | 0.4 |

[a] Minimum chi-square solution.

[b] Temperature range within confidence interval.

[c] Minimum emitting radius within confidence interval.

**Table 6**

**Measured Molecular Fluxes**

| Object | $H_2O$ | HCN | $C_2H_2$ | $CO_2$ | OH | $H_2$ S(1) |
|---|---|---|---|---|---|---|
| AA Tau | 4.8 (0.4) | 6.6 (0.2) | 1.5 (0.2) | 1.3 (0.1) | 0.6 (0.1) | 0.30 (0.06) |
| BP Tau | 6.6 (0.4) | 5.4 (0.2) | <0.5 | 0.9 (0.1) | 0.8 (0.1) | 0.31 (0.09) |
| DG Tau | <2.6 | <2.8 | <1.9 | <1.7 | 6.3 (0.7) | <0.70 |
| DK Tau | 15.4 (0.9) | 5.9 (0.5) | 0.5 (0.2) | 1.4 (0.2) | 2.0 (0.3) | <0.20 |
| DO Tau | 4.9 (0.8) | <1.4 | <1.0 | 1.3 (0.4) | 4.1 (0.4) | <0.38 |
| GI Tau | 6.0 (0.7) | 7.0 (0.5) | 2.2 (0.2) | 0.7 (0.1) | 1.0 (0.1) | 0.59 (0.09) |
| GK Tau | 4.6 (0.8) | <0.6 | <0.5 | 0.4 (0.1) | 1.3 (0.1) | 0.35 (0.09) |
| RW Aur | 21.7 (2.1) | 9.6 (0.8) | 6.7 (0.5) | 6.8 (0.4) | 2.8 (0.5) | 1.31 (0.25) |
| UY Aur | 12.1 (1.4) | :3.7 (2.0) | <0.7 (0.3) | 4.5 (0.4) | 3.1 (0.4) | <0.59 |
| AS353A | <0.4 | <0.5 | <0.4 | 1.7 (0.1) | 1.0 (0.1) | 0.48 (0.09) |
| V1331Cyg | <1.0 | <0.6 | <0.5 | 0.5 (0.1) | 0.9 (0.1) | <0.21 |

**Notes**. Fluxes are $10^{-17}$ W m$^{-2}$. Upper limits are 2 sigma.

**Table 7**

**Measured Atomic Fluxes**

| Object | [NeII][a] | H I[b] | [FeII][c] |
|---|---|---|---|
| AA Tau | 1.25 (0.08) | 0.20 (0.06) | … |
| BP Tau | 0.28 (0.05) | 0.90 (0.05) | … |
| DG Tau | 25.7 (0.3) | 7.4 (0.05) | 6.4 (0.8) |
| DK Tau | 0.73 (0.13) | 1.4 (0.1) | … |
| DO Tau | 0.91 (0.18) | 2.7 (0.4) | … |
| GI Tau | 0.73 (0.08) | 1.0 (0.1) | … |
| GK Tau | 0.53 (0.07) | 1.1 (0.1) | … |
| RW Aur | 0.5: (0.2) | 5.3 (0.4) | … |
| UY Aur | 2.41 (0.44) | 1.5 (0.3) | … |
| AS353A | 1.10 (0.11) | 3.1 (0.1) | 3.7 (0.2) |
| V1331CYg | <0.2 | 2.2 (0.2) | 8.3 (0.3) |

**Notes.** Fluxes are $10^{-17}$ W m$^{-2}$. Upper limits are 2 sigma.

a The 12.81 µm [Ne II] line.

b The 12.37 µm H I line.

c The 17.93 µm [Fe II] line is coincident with a H$_2$O feature and cannot be measured in objects with water emission.